\pdfoutput=1
\documentclass[11pt]{article}

\usepackage{EMNLP2022}

\usepackage{times}
\usepackage{latexsym}
\usepackage{hyperref}
\usepackage{url}
\usepackage{amsmath}
\usepackage{mathtools}
\usepackage{amssymb}
\usepackage{booktabs}
\usepackage{adjustbox}
\usepackage{graphicx}
\usepackage{placeins}
\usepackage{hhline}
\usepackage{caption}
\usepackage{multirow}
\usepackage{graphicx}
\usepackage{booktabs}
\usepackage{todonotes}
\usepackage{placeins}
\usepackage{changepage}

\graphicspath{{figures/}} %Setting the graphics path

\definecolor{maroon}{RGB}{191, 96, 96}

\definecolor{purple}{RGB}{100, 0, 200}

\definecolor{mint}{rgb}{0.24, 0.71, 0.54}

\usepackage[T1]{fontenc}
\usepackage[utf8]{inputenc}

\usepackage{microtype}

\usepackage{booktabs,arydshln}
\usepackage{array}

\definecolor{maroon}{RGB}{191, 96, 96}

\definecolor{purple}{RGB}{100, 0, 200}

\definecolor{mint}{rgb}{0.24, 0.71, 0.54}

\newcolumntype{L}[1]{>{\raggedright\let\newline\\\arraybackslash\hspace{0pt}}m{#1}}
\newcolumntype{C}[1]{>{\centering\let\newline\\\arraybackslash\hspace{0pt}}m{#1}}
\newcolumntype{R}[1]{>{\raggedleft\let\newline\\\arraybackslash\hspace{0pt}}m{#1}}

\makeatletter
\def\adl@drawiv#1#2#3{%
        \hskip.5\tabcolsep
        \xleaders#3{#2.5\@tempdimb #1{1}#2.5\@tempdimb}%
                #2\z@ plus1fil minus1fil\relax
        \hskip.5\tabcolsep}
\newcommand{\cdashlinelr}[1]{%
  \noalign{\vskip\aboverulesep
           \global\let\@dashdrawstore\adl@draw
           \global\let\adl@draw\adl@drawiv}
  \cdashline{#1}
  \noalign{\global\let\adl@draw\@dashdrawstore
           \vskip\belowrulesep}}
\makeatother

\newcommand{\eg}{e.\,g. }
\newcommand{\ie}{i.\,e. }

\makeatletter
\newcommand{\captshort}{\let\blx@imc@ifciteseen\@firstoftwo}
\makeatother

\title{CODER: An efficient framework for improving retrieval through COntextual Document Embedding Reranking}

\author{George Zerveas$^1$,~~Navid Rekabsaz$^2$,~~Daniel Cohen$^{1,3}$,~~Carsten Eickhoff$^1$  \\
  \\
  $^1$AI Lab, Brown University, USA~~\texttt{\{george\_zerveas,carsten\}@brown.edu}\\
  $^2$Johannes Kepler University Linz, LIT AI Lab, Austria~~\texttt{navid.rekabsaz@jku.at}\\
  $^3$Dataminr, USA~~\texttt{daniel.cohen@dataminr.com}
}

\begin{document}

\maketitle

\begin{abstract}
%We show that they are all important ingredients for improving performance, 

Contrastive learning has been the dominant approach to training dense retrieval models. In this work, we investigate the impact of \emph{ranking context} -- an often overlooked aspect of learning dense retrieval models. In particular, we examine the effect of its constituent parts: jointly scoring a large number of negatives per query, using retrieved (query-specific) instead of random negatives, and a fully list-wise loss. To incorporate these factors into training, we introduce Contextual Document Embedding Reranking (CODER), a highly efficient retrieval framework. When reranking, it incurs only a negligible computational overhead on top of a first-stage method at run time ($\sim\!5$ ms delay per query), allowing it to be easily combined with any state-of-the-art dual encoder method. Models trained through CODER can also be used as stand-alone retrievers. Evaluating CODER in a large set of experiments on the MS~MARCO and TripClick collections, we show that the contextual reranking of precomputed document embeddings leads to a significant improvement in retrieval performance. This improvement becomes even more pronounced when more relevance information per query is available, shown in the TripClick collection, where we establish new state-of-the-art results by a large margin. %It thus acts as a lightweight performance enhancement framework for a wide range of dense retrieval models.
%while enabling the convenient inclusion of ranking constraints beyond relevance, e.g. related to fairness.
%It employs a list-wise loss and jointly scores a large set of retrieved candidate documents, rather than randomly sampled documents, for each query.
%, while potentially transforming the representation of each document in the context of the other candidates as well as the query itself.
%When scoring a document representation based on its similarity to a query, the model is thus aware of the representation of its ``peer'' documents within the same retrieval context. 
%We investigate the effect of training through contextual reranking of document embeddings and show that our approach leads to substantial improvement in retrieval performance over pair-wise scoring of candidate documents in isolation from one another. 
%Crucially, CODER incurs only a negligible computational overhead ($\sim\!5.5$ ms delay per query) on top of a first-stage method at run time, allowing it to be easily combined with any state-of-the-art dense retrieval method.
%\footnote{The code for CODER will be publicly shared on Github after the review process. It has been anonymously shared as a material on the review platform}
%enables additional interesting avenues for research in retrieval

\end{abstract}

\section{Introduction}
\vspace{-2mm}
% \subsection{Triplet-based contrastive learning and ``in-batch'' negatives}

Neural text retrieval models typically rely on a contrastive training optimization that uses \textit{(query, positive document, negative document)} triplets as training samples. This scheme is especially popular, as it is well-suited to the computational constraints of large transformer-based language models such as BERT~\cite{devlin_bert_2019} and its variants for retrieval~\cite{nogueira_passage_2020, khattab_colbert_2020, zhan_repbert_2020, zhan_learning_2020}. Referred to as pair-wise training, the model is asked to score the similarity between the query embedding and a ground truth relevant (positive) document embedding, higher than the one between the query and a negative document embedding.

While such contrastive learning approaches are effective, by only considering pairs of positive and negative documents at a time, they (1)~deviate from the target objective of comparing a query against many documents while discarding inter-document information, and (2)~depart from core list-wise evaluation metrics like nDCG~\citep{ranking_loss_calibration_neurips2012}.

Addressing the first shortcoming, recent works have employed ``in-batch'' negatives: given a batch containing \textit{(query, positive document)} tuples, the negatives to a tuple are set as the known positive documents from other queries within that batch. (e.g.~\citet{karpukhin_dense_2020, luan_sparse_2021, zhan_repbert_2020, hofstatter_efficiently_2021, qu_rocketqa_2021}). Although this approach efficently increases the number of negatives to improve performance, the presence of \textit{hard} negative samples\footnote{Hard negatives look similar in topic and term distribution to relevant documents, while not actually satisfying the information need. } is critical in achieving state-of-the-art (SOTA)~\cite{xiong_approximate_2020, zhan_optimizing_2021, qu_rocketqa_2021, hofstatter_efficiently_2021}. 
%To ensure this, an effective training regime needs to contain high-quality negative samples in sufficient quantity. As in-batch negatives cannot readily fulfill these criteria, this often results in significant computational requirements.

%While existing approaches do not account for the interactions between documents at retrieval time,
The rich literature on \emph{learning-to-rank (L2R)} has outlined compelling reasons for taking into account the context of other candidate documents being ranked for a query when scoring each document~\cite{cao_learning_2007,ai_learning_2019, ai_learning_2018}, realized in various \textit{list-wise} optimizations. Such approaches allow for the model to directly optimize IR metrics as opposed to a surrogate pair-wise loss as seen in the contrastive training regime. Despite these list-wise approaches achieving competitive results in a variety of ranking situations, considerations of computational complexity and stochastic stability practically relegate them to shallow neural models over handcrafted feature vectors~\citep{bruch_stochastic_l2r2020,pang_setrank_2020, Chen2021PoolRankMP}.

In this paper, we extend existing work in negative sampling and list-wise learning to allow for large pre-trained language models to take advantage of the context found in a large, \textit{coherent} set of candidate documents. %While recent work on transformer-based dense retrieval has indirectly discovered and partially explored aspects of the effect of context by casting it separately as a question of the overall number of negatives and of mining appropriate hard negatives, we expressly investigate the importance of context when training dense retrieval models. 
We particularly examine the effect of constituent parts of the query context, \ie (1)~jointly scoring a large number of negatives, (2)~using retrieved (query-specific) instead of random negatives, and (3)~a fully list-wise loss. To this end, we introduce \emph{COntextual Document Embedding Reranking (CODER)}, a highly efficient and generic fine-tuning framework that for the first time enables incorporating context, previously only considered in learning-to-rank neural networks, into transformer-based language models used in state-of-the-art dense retrieval. CODER acts as a lightweight performance enhancing framework that operates on precomputed document embeddings, while transforming the query to account for new list-wise context information over a large number of query-specific hard negative candidate documents. It can be applied to virtually any existing dual-encoder model, and used both for single- as well as two-stage dense retrieval.

Our contribution is three-fold: (1)~We introduce an efficient framework which enables leveraging ranking context; (2)~We conduct a large set of experiments on the MS~MARCO~\cite{bajaj_ms_2018} and TripClick~\cite{rekabsaz_tripclick_2021} collections and show that CODER can considerably enhance the effectiveness of a wide class of dense retrieval models at minimal computational cost, while achieving new SOTA results on TripClick; (3)~We explore the impact of the constituent parts of ranking context in learning effective models. Our code and trained resources are available in \url{https://github.com/gzerveas/CODER}.

\section{Related Work}
\label{sec:related}

% The CODER framework addresses a dilemma in state-of-the-art IR: on the one end are powerful but slow transformer models that model direct interactions between query and document terms through attention~\cite{nogueira_passage_2020, khattab_colbert_2020}. They can only be used as rerankers within a cascade system due to their cost. On the other end are fast, but less effective dense retrieval models, which employ a dual transformer encoder architecture to separately encode the query and document sequences~\cite{karpukhin_dense_2020, luan_sparse_2021, zhan_repbert_2020, xiong_approximate_2020, qu_rocketqa_2021, hofstatter_efficiently_2021}, and for inference rely on the efficient computation of the dot product through high-performing Approximate Nearest Neighbors libraries such as FAISS~\cite{johnson_billion-scale_2017} to evaluate the similarity between extracted query and document representations. CODER improves the performance of these existing (base) dense retrieval dual encoder models and can be used either for rapid reranking in a cascade system, incurring minimal delay even in the case of full collection retrieval.

Recent work has demonstrated the importance of the quality of negative documents used during fine-tuning. \citet{xiong_approximate_2020} periodically re-encode every query and document in the collection during training in order to mine the most difficult documents to use as negative candidates via approximate nearest neighbor (ANN) search. Improving on this slow and resource-intense process, \citet{zhan_learning_2020} (published as \citet{zhan_optimizing_2021}) forego fine-tuning of the document encoder, instead only fine-tuning the query encoder while dynamically mining negatives. TAS-B~\cite{hofstatter_efficiently_2021} also improves the quality of negatives by clustering semantically similar queries, such that the in-batch negatives are indirectly related to the ground truth document. While we contribute to this line of research, we show that one can avoid such complexity and directly benefit from list-wise optimization applied on a large, coherent and informative context of candidate documents, retrieved for each query in advance.%, created from the candidates retrieved for the same query by a competent retrieval method. 

%Given the fixed document representations, only the query needs to be dynamically encoded to identify challenging candidate documents via ANN.
%Importantly, while the authors argue that fixed negative candidates will always under-perform compared to the above dynamic approaches, we demonstrate that using a large number of even fixed negative candidates in a list-wise setting can outperform these approaches.%, although we believe that both approaches can ideally be used in combination.

% \begin{figure}[t]
%     \centering
%     \includegraphics[scale=0.35]{coder_diagram.png}
%     \caption{Schematic diagram of the CODER method. The architecture consists of two main components: the query encoding module, which embeds a tokenized query, and the document set scoring module, which jointly scores a set of $N$ precomputed candidate document embeddings. Here, we experiment specifically with $\varphi = \mathbf{X} \cdot g\left(\mathbf{Z} \right)$, combined with a list-wise loss and large $N=1000$, which we show allows to effectively leverage ranking context.}
%     \label{fig:generic_schematic_diagram}
%     \vspace{-0.4cm}
% \end{figure}

Moving from quality to quantity of negative candidates, RocketQA~\cite{qu_rocketqa_2021} drastically increases the quantity of random in-batch negatives to several thousands by sharing negatives across at least 8 V100 GPU instances. Despite its huge size, this pool of sampled negatives includes only 4 retrieved hard negatives per query and is otherwise almost entirely random, with no shared context across documents. Unlike CODER, this approach necessitates training an expensive cross-encoder model to ``denoise'' (filter out) retrieved candidates, otherwise yielding poor performance. Moreover, it involves using the cross-encoder to pseudo-label additional data samples, an approach adopted by the currently top performing dual encoder models following up this work, RocketQAv2~\cite{ren_rocketqav2_2021}, which additionally leverages list-wise cross-encoder teaching, and PAIR~\cite{ren_pair_2021}, which includes a loss term capturing similarity between passages.

To address the increasing complexity and computational requirements of training pipelines, \citet{Gao2021CondenserAP, Gao2021UnsupervisedCA} instead propose corpus-specific, self-supervised pre-training with a bespoke transformer add-on (see Baselines in Section \ref{sec:exp_setup}). An orthogonal approach to reduce computational requirements focuses on jointly optimizing query optimizer and product quantization for ANN search~\cite{zhan2021jointly}.

List-wise loss functions have been extensively used within learning-to-rank (L2R), although they have been limited to either shallow neural models~\cite{bruch_stochastic_l2r2020,cao_learning_2007} or various deep networks~\citet{pasumarthi_tf-ranking_2019, ai_learning_2018, ai_learning_2019, pang_setrank_2020} in works focusing on L2R datasets. These consist of handcrafted feature vectors representing query-document similarity such as term overlap, click-through rate, BM25 scores, and other salient features.
Effectively applying L2R concepts to transformer-based language models used for ad-hoc dense retrieval is a non-trivial challenge, and represents CODER's extension of prior works.
%We note that \citet{cohen_td_ir2021} demonstrated that certain RL algorithms can allow for a list-aware model, while the combination of RL and deep neural models is exceedingly expensive and relatively unstable.

Finally, there is a body of work utilizing large pre-trained language models for retrieval in cross-encoder~\cite{nogueira_passage_2020,lin2021pretrained,hofstatter_establishing_2022}, late cross-encoder~\cite{khattab_colbert_2020, santhanam_colbertv2_2021}, and generative rankers~\cite{10.1145/3471158.3472229}, as well as query/document expansion and indexing~\cite{zheng2020bert,naseri2021ceqe,mallia2021learning,nogueira2020document,gao2021coil}. Co-BERT~\cite{chen_incorporating_2022}, a recent method leveraging ranking context,  uses a cross-encoder BERT Reranker to select candidates and to compute feature vectors as input for the L2R methods above~\cite{pang_setrank_2020} through query-document term interactions. In comparison, CODER is orders of magnitude more efficient both during training and inference.

All existing approaches either advocate for using a handful of hard negatives (e.g.~\cite{karpukhin_dense_2020}), or compromise with it due to computational constraints. By recognizing the importance of context, our proposed framework is the first to allow the combination of quantity and quality of negatives for training SOTA dual encoder models with very modest computational resources.

\vspace{-0.2cm}
\section{Method} \label{sec:method}
\begin{figure}[t]
    \centering
    \includegraphics[scale=0.5]{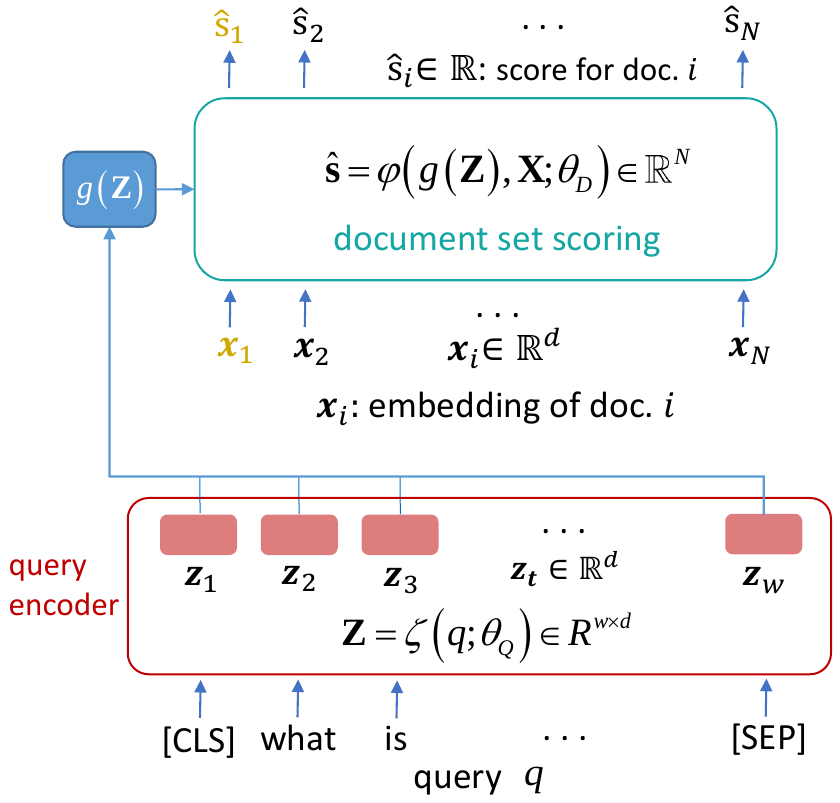}
    \caption{Schematic diagram of the CODER method. The architecture consists of two main components: the query encoding module, which embeds a tokenized query, and the document set scoring module, which jointly scores a set of $N$ precomputed candidate document embeddings. Here, we experiment specifically with $\varphi = \mathbf{X} \cdot g\left(\mathbf{Z} \right)$, combined with a list-wise loss and large $N=1000$, which we show allows to effectively leverage ranking context.}
    \label{fig:generic_schematic_diagram}
    \vspace{-0.4cm}
\end{figure}

\vspace{-0.1cm}
CODER involves fine-tuning a pre-trained query encoder to learn a query representation that is as proximal as possible to the representation of the ground-truth relevant document(s), by adjusting it to better account for the context of multiple query-related documents. The architecture consists of two main components (Fig.~\ref{fig:generic_schematic_diagram}): a query encoder, which builds a query representation, and the document set scoring module, which, given a query representation, jointly scores a set of $N$ precomputed embeddings of positives and hard negatives retrieved by an arbitrary retrieval method, $M_C$. Using precomputed document embeddings reduces computational costs (memory, FLOPs) by a factor of $N$. Thus, unlike all existing approaches, we can afford to use a large number of such hard negatives ($N=1000$ in our experiments, unless otherwise noted).

\vspace{-0.2cm}
\subsection{Architecture}
\vspace{-0.1cm}
\subsubsection{Query Encoder}

The query encoder can be any pre-trained transformer encoder, such as BERT~\cite{devlin_bert_2019}, DistilBERT~\cite{sanh_distilbert_2020}, RoBERTa~\cite{liu_roberta_2019}, or ERNIE~\cite{zhang_ernie_2019}. We initialize its weights from an existing model already fine-tuned for retrieval, which we call ``\textit{base model}'', $M_D$.

Formally, for each query token $q_t$, $t\in \mathbb{N}: 1\leq t \leq w$, where $w$ is the length of the tokenized query sequence, it extracts a vector representation $\mathbf{z}_t \in \mathbb{R}^d$, where $d$ is the encoder's internal representation dimension. These vectors can be linearly projected to a space of different dimensionality and become  $\mathbf{z}_{t}' \in \mathbb{R}^{d'}$, to match the dimensionality of the document embeddings $d'$, in case the latter differs. In the general case we thus denote the extracted representation of a query $q$ as: 
\begin{equation}
    \mathbf{Z'} = [\mathbf{z}_{1}'; \dots; \mathbf{z}_{w}'] = \zeta \left(q; \theta_Q \right) \in \mathbb{R}^{w \times d'},
\end{equation}
where $\mathbf{\theta}_Q$ are the parameters of the query encoder. The individual token representations are aggregated into a single vector using an aggregation function $g$. In our experiments, we let $g(\mathbf{Z'}) = \frac{1}{w}\sum_t \mathbf{z'}_t$ be the mean when using RepBERT~\cite{zhan_repbert_2020}, and $g(\mathbf{Z'}) = \mathbf{z'}_1$ (i.e. the representation of the \texttt{[CLS]} token) when using TAS-B~\cite{hofstatter_efficiently_2021} as the base model for implementing the query encoder $\zeta$ (see Section~\ref{sec:exp_setup}).
%, which for example can compute the mean over token positions $t$, or select the vector corresponding to $t=1$ (i.e. the \texttt{[CLS]} token in BERT-based models), such that $g(\mathbf{Z'}) \in \mathbb{R}^{d'}$, but can alternatively be the identity function.

% \vspace{-0.2cm}
\subsubsection{Document scoring function}
The document set scoring function is represented by $\varphi$ which produces a set of $N$ scalar relevance scores $\hat{s}_i \in \mathbb{R}$, $i\in \mathbb{N}: 1\leq i \leq N$. It takes as an input the aggregated query representation $g(\mathbf{Z'})$ from the previous section, and a set of $N$ document embeddings $\mathbf{x}_i \in \mathbb{R}^m$, $i\in \mathbb{N}: 1\leq i \leq N$, precomputed by the base model.
%, which correspond to the top $N$ candidates for the query $q$ retrieved by an arbitrary retrieval method. 
Using learnable linear projections, the dimensionality of the document embeddings can potentially be changed to accommodate different scoring functions (e.g., transformer blocks with an internal representation dimension $d' \neq m$), while matching the dimensionality of the query embeddings. Succinctly, the output relevance scores are:
\vspace{-1mm}
\begin{equation}
    \mathbf{\hat{s}} = \varphi \left(g(\mathbf{Z'}), \mathbf{X} ; \mathbf{\theta}_D\right) = \mathbf{X} \cdot g\left(\mathbf{Z} \right) \in \mathbb{R}^{N},\label{eq:predicted_scores}
    %\vspace{-1mm}
\end{equation}
where $\mathbf{X} = [\mathbf{x}_1; \dots; \mathbf{x}_N]\in \mathbb{R}^{N \times m}$ are the $N$ document embeddings, $\theta_D$ are the parameters of the scoring function, and $d' \equiv d$ (i.e. query and document embeddings have the same dimensionality).

While a variety of functions can be used as a scoring function in our framework, including transformers (see Appendix Section~\ref{sec:transformer_document_scoring}), for all results presented in this work we leverage the simple inner product, which interestingly achieves significant performance improvements even without the contextualized transformation of document embeddings. It thus appears that jointly scoring a large number of query-specific candidates for the same query within a list-wise loss establishes a strong enough context for improving performance (see Section~\ref{sec:context_without_doc_relationships}). Beyond computational efficiency, the main advantage of the above function is that it facilitates directly using the fine-tuned query encoder for dense retrieval (single-stage) through fast approximate nearest neighbor search. Instead, a non-linear scoring module would only allow using the model for reranking in a two-stage retrieval setting (candidate retrieval, followed by reranking).

% \vspace{-0.2cm}
\subsection{Training through CODER}

The document representations $\mathbf{X}$ are precomputed using the \textit{document} encoder part of any state-of-the-art dual encoder retrieval model $M_D$. To accelerate training convergence, we initialize our query encoder $\zeta$ from the query encoder of the same dual encoder retrieval model. Throughout training, the parameters $\mathbf{\theta}_Q$ of the query encoder (and $\mathbf{\theta}_D$ of the scoring function $\varphi$, if the latter is learnable) are fine-tuned. To support a memory and compute-efficient training setting without the need for large or parallel GPUs, the document representations $\mathbf{X}$ remain fixed, although in general they can be transformed by function $\varphi$ before computing the document similarity scores. As with all dense retrieval methods (e.g., \cite{zhan_repbert_2020, xiong_approximate_2020, zhan_optimizing_2021, hofstatter_efficiently_2021}), document representations are assumed to have been precomputed and indexed for fast inference runtimes. 

A key difference between CODER and all dense retrieval methods is that, for each query, along with the $k$ positive (ground-truth) documents, the model is trained to jointly score a \textit{large number} $N-k$ of top candidate documents retrieved by some \textit{candidate} retrieval method $M_C$.  %During training, in case $M_C$ failed to retrieve the ground-truth document(s) corresponding to a query within the set of $N$ candidates, we replace the bottom-ranked candidate(s) with the ground-truth document(s). 
We note that $M_C$ does not need to be the same method as the base method that provides the document representations and the query encoder. This potentially allows leveraging methods with different characteristics (\eg methods with higher recall versus precision, or a lexical overlap / sparse representation such as BM25), and can prove beneficial, as shown in Section~\ref{sec:results}.

\vspace{-3mm}
\subsection{Loss function}\label{sec:loss_function}
\vspace{-0.1cm}

To best take advantage of jointly scoring $N$ documents for each query, we choose the ListNet loss~\cite{cao_learning_2007}, which is the KL-divergence between a distribution over the predicted scores $\mathbf{\hat s}$ (given by Eq.~\eqref{eq:predicted_scores}) for the $N$ candidate documents, and a distribution over the target (ground-truth) relevance labels $\mathbf{y}\in \mathbb{R}^{N}$, given by the dataset for the same set of candidates (the relevance score of positive documents is a positive scalar, while for negative or documents whose label is not explicitly defined it is set to $-\infty$):

\vspace{-5mm}
\begin{equation}
\begin{split}
\mathcal{L}\left( \mathbf{y},{\bf{\hat s}} \right) & = {\mathop{\rm D_{KL}}\nolimits} \left( {\sigma (\mathbf{y})\mid \mid{\sigma(\bf{\hat s})}} \right) \\ 
& = - \sum\limits_{i = 1}^N {\sigma{{({\bf{y}})}_i}} \log \frac{\sigma(\mathbf{\hat s})_i}{{\sigma{{({\bf{y}})}_i}}}\label{eq:loss}
\end{split}
\vspace{-2mm}
\end{equation}
where $\sigma$ denotes the softmax function. 
Jointly scoring a large number of \textit{retrieved} candidates for each query, in combination with the KL-divergence loss, distinguishes our method from existing dense retrieval methods. This combination establishes and exploits a \textit{context} for each query and it is key for obtaining a performance improvement over the base method, as we show in Section~\ref{sec:ablation} (and further discuss in Appendix Section~\ref{sec:context_without_doc_relationships}). The benefit is expected to be even greater for datasets which include multiple document labels per query, optionally defined over several levels of relevance. We show such results in Section~\ref{sec:tripclick_results}.

\section{Experimental Setup}\label{sec:exp_setup}

\paragraph{Datasets.}
We conduct the experiments on passage and document retrieval tasks,\footnote{When referring to the unit of retrieval we use the terms ``passage'' and ``document'' interchangeably.} using two large publicly available IR collections in the domains of web and health retrieval. The first dataset is MS~MARCO Passage Retrieval dataset~\cite{bajaj_ms_2018} used for training and evaluation. We evaluate the models trained on MS~MARCO also on TREC Deep Learning Passage Retrieval track 2019 and 2020~\cite{craswell2020overview,craswell2021overview}. The second dataset is TripClick, a recently introduced health document retrieval dataset~\cite{rekabsaz_tripclick_2021} used for training and evaluation. Details of the collections are provided in Section~\ref{sec:data_appendix} of the Appendix. While the training data of MS~MARCO contains only approximately 1 relevance judgement per query, the TripClick collection has the advantage of providing a much larger set of relevance information, namely approximately 42, 9, and 3 data points per query in HEAD, TORSO, and TAIL sets, respectively. As shown in the next section, this is particularly beneficial when optimizing over a large ranking context in a list-wise manner. Evaluation details are given in Section~\ref{sec:evaluation} of the Appendix.

\paragraph{Baselines.}%\label{sec:baselines}

%To evaluate the effectiveness and generality of our method, 
We choose several dense retrieval models as baselines, \ie ``base models'' subjected to CODER fine-tuning:\\
1.~RepBERT~\cite{zhan_repbert_2020}, a BERT-based model with a typical dual encoder architecture which underpins all state-of-the-art dense retrieval methods, trained using a triplet Max-Margin loss.\\
2.~TAS-B~\cite{hofstatter_efficiently_2021}, which, besides being a top-performing dense retrieval method on the MS MARCO~/~TREC-DL 2019, 2020 datasets, it also represents methods that have been optimized with respect to their training process (details in Section~\ref{sec:related}).\\
% In particular, TAS-B involves a sophisticated selection of negative documents through clustering of topically related queries, as well as knowledge distillation from two powerful, heavyweight cross-encoder (term-interaction) models, BERT-Reranker~\cite{nogueira_passage_2020} and ColBERT~\cite{khattab_colbert_2020}.
3.~Finally, to explore the limits of CODER, we use it to fine-tune a trained CoCondenser retriever~\cite{Gao2021UnsupervisedCA}, the state-of-the-art dense retrieval model \textit{that does not} make use of query-document term interactions, cross-encoder teacher models or additional pseudo-labeled data samples, but instead relies on extensive corpus-specific, self-supervised pre-training using a special architecture and contrastive loss component. It is a particularly challenging baseline for our CODER framework, because it has been trained through  (a)~mining for hard negatives using a trained version of the model itself, and (b)~the Negative LogLikelihood (InfoNCE) loss, which is “nearly” list-wise (it differs from our KL-divergence loss only when there are more than one positive candidates).
%For completion and to have a general performance comparison with some cross-encoder rankers, we also include published metrics for relevant related work for comparison on MS MARCO and TREC DL tracks: ColBERT and L2Re (ANCE)~\cite{zhan_learning_2020}.
%Lastly, we include published metrics for relevant related work for comparison on MS MARCO and TREC DL tracks: ColBERT and L2Re (ANCE)~\cite{zhan_learning_2020}.
%We use the same batch size (32) as \citet{hofstatter_efficiently_2021}, and fine-tune the fully trained TAS-B query encoder from \citet{hofstatter_efficiently_2021}. 

%We believe that achieving performance improvements over these baselines demonstrates the broad applicability of our approach on a wide class of the state-of-the-art retrieval models and a very promising potential for conferring performance benefits even on models with an optimized training process. All baselines are extensively fine-tuned on the target collection using a validation set to determine stopping criteria.

\begin{table*}
\centering
\resizebox{0.96\textwidth}{!}{%
{\setlength\doublerulesep{0.7pt}
\begin{tabular}{@{}l|ll|ll|ll|c@{}}
\toprule
                    \multirow{2}{*}{\textbf{Model}}        & \multicolumn{2}{c|}{\textbf{MS MARCO dev}} & \multicolumn{2}{c|}{\textbf{TREC DL 2019}}             & \multicolumn{2}{c|}{\textbf{TREC DL 2020}}           &  \multirow{2}{*}{\textbf{\begin{tabular}[c]{@{}c@{}}Latency\\ (ms/query)\end{tabular}}}                                                                     \\
             & \textbf{MRR@10}     & \textbf{nDCG@10}     & \textbf{MRR@10}          & \textbf{nDCG@10}         & \textbf{MRR@10}        & \textbf{nDCG@10}         &  \\ \midrule
BM25 {[}Anserini{]}               & 0.187               & 0.234                & 0.843 / 0.682           & 0.497 / 0.417            & 0.820 / 0.655          & 0.488 / 0.412            & $50^1$                                                      \\
L2Re(ANCE)~\cite{zhan_learning_2020}                  & 0.341               &          -            &    -                     & 0.675                   &           -             &            -             & 47                                                                    \\
Co-BERT$^*$~\cite{chen_incorporating_2022} & - & - & 0.958  &  0.700 & 0.839 & 0.699  & > 1000 \\
$^2$ColBERT$^*$v1; v2                     & 0.360;~~0.397                &          -            &             -             &               -           &            -            &               -           & 458                                                                   \\
$^2$BM25 $\rightarrow$ ColBERT$^*$              & 0.349               & -                    & -                        & -                        & -                      & -                        & {[}BM25{]} + 61                                                       \\
$^3$RocketQAv1; v2  &  0.370;~~0.388  &  -  & - & - & - & - & - \\
CoCondenser [our evaluation] &	0.381  &	0.446  &	0.971 / 0.879  &	0.715 / 0.656  & 	0.937 / 0.833  &	0.680 / 0.618 &	$\simeq$ [RepBERT] \\

 \midrule\midrule\midrule
RepBERT~(abbrev: RB) [our eval.]                          & 0.304               & 0.359                & 0.917 / 0.766           & 0.616 / 0.548            & 0.902 / 0.763          & 0.621 / 0.561            & 70                                                                   \\
BM25 $\rightarrow$ RepBERT                    & 0.317               & 0.373                & 0.969   / 0.795         & 0.674   / 0.593          & 0.893   / 0.781        & 0.640   / 0.579          & {[}BM25{]} + 5.8                                                      \\ \cdashlinelr{1-8}%\midrule
BM25 $\rightarrow$ CODER(RB, BM25)       & 0.326               & 0.384                & 0.953 / 0.798           & 0.675 / 0.600            & 0.914 / \textbf{0.816} & 0.654 / 0.593            & {[}BM25{]} + 5.8                                                      \\
BM25 $\rightarrow$ CODER(RB, RB)    & \textbf{0.327}$^{ab}$      & \textbf{0.385}$^{ab}$       & 0.953  / \textbf{0.806} & 0.677   / \textbf{0.603$^{a}$} & 0.898   / 0.787        & 0.672   / \textbf{0.611}$^{ab}$ & {[}BM25{]} + 5.8                                                      \\
RB $\rightarrow$ CODER(RB, RB) & 0.324               & 0.383                & 0.905   /  0.785        & 0.650   / 0.593          & 0.918 / 0.785          & 0.660   / 0.598          & {[}RepBERT{]} + 5.8                                                   \\
CODER(RB, BM25)              & 0.311               & 0.368                & 0.855   / 0.750         & 0.606   / 0.552          & 0.906   / 0.790        & 0.603   / 0.550          & {[}RepBERT{]}                                                         \\
CODER(RB, RB)           & 0.325               & 0.384                & 0.905   / 0.785         & 0.652   / 0.593          & 0.918   / 0.785        & 0.660   / 0.598          & {[}RepBERT{]}                                                         \\
\midrule%\midrule
TAS-B~\cite{hofstatter_efficiently_2021}                       & 0.340               & 0.402                & 0.892                    & 0.712                    & 0.843                  & 0.693                    & 64                                                                    \\
TAS-B {[}our evaluation{]}  & 0.344               & 0.408                & 0.951   / 0.875 & 0.721   / 0.659          & 0.921   / 0.832        & 0.685   / 0.620          & <~50                                                          \\
BM25 $\rightarrow$ TAS-B                & 0.343               & 0.404                & 0.971   / 0.857          & 0.723   / 0.648          & 0.918   / 0.838        & 0.696   / \textbf{0.633} & {[}BM25{]} + 5.5                                                      \\ \cdashlinelr{1-8}%\midrule
BM25 $\rightarrow$ CODER(TAS-B, BM25)   & 0.349      & 0.409       & 0.983 / 0.872   & 0.727 / 0.654            & 0.935 / \textbf{0.846}          & 0.690 / 0.629            & {[}BM25{]} + 5.5                                                      \\
BM25 $\rightarrow$ CODER(TAS-B, TAS-B)  & 0.350               & 0.411                & 0.971   / 0.828          & 0.728   / 0.654          & 0.926   / \textbf{0.846}        & 0.693   / \textbf{0.630} & {[}BM25{]} + 5.5                                                      \\
TAS-B $\rightarrow$ CODER(TAS-B, TAS-B) & \textbf{0.355}$^{ab}$      & \textbf{0.419}$^{ab}$       & 0.966   / 0.857          & 0.728   / \textbf{0.668} & 0.923   / \textbf{0.844}        & 0.686   /  0.623         & {[}TAS-B{]} + 5.5                                                     \\
CODER(TAS-B, BM25)          & 0.347               & 0.409                &  0.965 / \textbf{0.890}           & 0.723 / 0.665            & 0.934 / 0.835 & 0.678 / 0.612            & {[}TAS-B{]}                                                           \\
CODER(TAS-B, TAS-B)         & \textbf{0.355}$^{ab}$      & \textbf{0.419}$^{ab}$       & 0.966   / 0.857          & 0.728   / \textbf{0.668} & 0.923   / \textbf{0.844}        & 0.686   / 0.623          & {[}TAS-B{]}                                                           \\ \bottomrule
\end{tabular}
}
}
\caption{Performance for passage ranking when applying CODER to the RepBERT (middle section) and TAS-B (bottom section) base methods. In the notation $M_F \rightarrow \text{METHOD}(M_D, M_C)$: $M_F$ is the method used for first stage retrieval when using METHOD for reranking, $M_D$ is the base method and $M_C$ is the retrieval method which provides the context (candidate) passages during training.
\textbf{Bold font denotes best results within the same section (separated by continuous rules)}. Results of the statistical significance tests (paired $t$-test) are reported only for the best performing models, where the symbols $^a$ and $^b$ denote a significant improvement ($p<0.05$) with respect to the base $M_D$ and BM25$\rightarrow M_D$, respectively.
For TREC DL, two values are given for each metric separated by a slash, corresponding to the lenient / strict (official) interpretation of relevance labels. Models with $^*$ use cross-encoder term interactions. Rows with $^2$ are  from \citet{khattab_colbert_2020, santhanam_colbertv2_2021}, and with $^3$ from ~\cite{qu_rocketqa_2021, ren_rocketqav2_2021}.}
\label{tab:CODER_MSMARCO}
\vspace{-4mm}
\end{table*}

%As discussed in Section~\ref{sec:method}, there are a variety of possible configurations when using the CODER framework for retrieval. To denote the different configurations, 
\vspace{-0.1cm}
\paragraph{Configurations.} For the CODER framework, we use the following notation: 
$M_F~\rightarrow~\text{CODER}(M_D, M_C)$, 
where $M_D$ is the base model used to encode documents into document embedding vectors (and initialize the query encoder weights), $M_C$ is the first-stage retrieval method used to procure the candidate (context) documents reranked during the CODER training process, and $M_F$ is the retrieval method used as a first stage when CODER is used as a reranking method during inference; $M_F$ vanishes in case CODER is used directly for single-stage dense retrieval, and the notation $\text{CODER}(M_D, M_C)$ is used instead. In addition to TAS-B and RepBERT, we also experiment with BM25 (Anserini implementation~\cite{yang_anserini_2017}) as $M_C$ and $M_F$ methods. Specific hyperparameter details can be found in Table~\ref{tab:coder_config} in Appendix.

\vspace{-0.2cm}
\section{Results}\label{sec:results}

\vspace{-0.1cm}
\subsection{Results on MS MARCO and TREC DL}\label{sec:msmarco}

%Starting from a RepBERT model which has been trained until reaching the peak of its performance~\cite{zhan_repbert_2020}, we fine-tune it for 87k training steps (batches of 32 queries, each with 1000 candidates), which took about 4.5 hours on a single NVIDIA TITAN RTX GPU - further fine-tuning did not improve MRR on our validation set.

After only a fast and efficient fine-tuning (3.5 hours for TAS-B, 4.5 hours for RepBERT on
a single NVIDIA TITAN RTX GPU), we observe a substantial performance benefit when applying our CODER framework to TAS-B and RepBERT, as seen in Table~\ref{tab:CODER_MSMARCO}. 

CODER improves retrieval performance remarkably compared to both the original (single-stage) RepBERT, as well as two-stage cascade BM25$\rightarrow$RepBERT. CODER confers the largest performance benefit on RepBERT when reranking BM25 candidates in a cascade. The single-stage retriever fine-tuned through CODER is much improved compared to the original RepBERT, and almost as effective as the cascade, without introducing any latency or complexity.
%An interesting result is the performance difference of RepBERT and BM25$\rightarrow$RepBERT. While RepBERT was proposed and trained as a first-stage retrieval model, it gains a substantial and consistent improvement when reranking BM25 candidates despite RepBERT being an overall much stronger retrieval method than BM25, with a much better recall (see Appendix~\ref{sec:repbert_vs_bm25} for possible explanations).
%In any case, CODER performs close to its maximum performance regardless of whether it is reranking BM25 candidates or directly ranking the entire collection.

Our framework also significantly improves the performance of the highly optimized TAS-B method, which leverages hard negatives and dual knowledge distillation from two powerful cross-encoder models, BERT-Reranker and ColBERT. We furthermore observe that using the CODER-trained query encoder directly for single-stage dense retrieval is exactly as effective as using CODER to rerank TAS-B candidates in a cascade, on both MS MARCO and TREC DL tracks. This result suggests that under certain conditions, CODER can generalize its ranking function from the provided training context (limited set of fixed hard negatives) to the entire dataset, even without the use of dynamic negative mining, huge batch sizes or ``denoising'' as seen in previous work~\citep{qu_rocketqa_2021, xiong_approximate_2020}.

Putting these results into context, we can see that simply by training through contextual reranking, a dual encoder model such as TAS-B can improve to the point that its reranking effectiveness is higher than a powerful model such as ColBERT, a term-interaction model still fast enough to be practically considered for real-time reranking, but at a fraction of the reranking latency cost (5.5 ms vs 61 ms, i.e., less than 1/10th). Its single-stage ranking performance  approaches the one by ColBERT, while being about 10x faster (about 50 ms vs.\ 458 ms), making it a top-performing method within its latency class.

%We also note that our framework is compatible with, and can incorporate techniques proposed in other work, such as dynamic retrieval of negatives from \citet{zhan_learning_2020} and ``denoising'' negatives from \citet{qu_rocketqa_2021}, to potentially further improve performance.
%At the same time, as mentioned in Section~\ref{sec:fairness_calibration}, this framework enables fairness regularization and better score calibration.
%170 ms, 5.3 ms per query in another measurement - depends on GPU

Finally, to test the limits of CODER (see Section~\ref{sec:exp_setup}), we apply it to fine-tuning a trained CoCondenser retriever, the SOTA dense retrieval model that does not rely on using a cross-encoder in its pipeline. We observe a slight improvement of 0.002 MRR@10 and 0.004 Recall@10 (the latter statistically significant) on the MS MARCO validation set. This smaller improvement on MS MARCO is expected, given that it is a dataset providing very few relevance judgements per query and thus (a) poor context for training, and (b) an evaluation setting that may be unsuitable to resolve differences in ranking effectiveness for a contextually-trained model. For this reason, we additionally evaluate the models on TripClick.

\begin{table}[t]
\centering
\resizebox{0.46\textwidth}{!}{%
%{\setlength\doublerulesep{0.5pt}
\begin{tabular}{@{}l ll}
\toprule
%\multicolumn{1}{c|}{\textbf{TripClick TEST}} 
\textbf{Model} & \textbf{MRR@10}   & \textbf{nDCG@10}\\ \midrule
BM25$^1$                     & 0.276         & 0.224 \\
Transformer-Kernel$^1$                  & 0.434         & 0.284                  \\ 
BERT-Dot (SciBERT)$^2$                  & 0.530          & 0.243                   \\
BERT-Cat (SciBERT; PMBERT)$^2$                  & 0.595; 0.582         & 0.294; 0.298         \\
%BERT-Cat(PubMedBERT-Full)$^2$                  &  0.582         & 0.298           & -  \\
%BERT-Cat(PubMedBERT-Abs.)$^2$                  &   0.587        & 0.296           & -  \\
\midrule \midrule %\midrule
RepBERT  (abbrev: RB)                       & 0.526          & 0.255          \\
BM25 $\rightarrow$ RepBERT                      & 0.538        & 0.262     \\ \cdashlinelr{1-3}%\midrule
RB $\rightarrow$ CODER(RB, RB)         & \textbf{0.637}* & \textbf{0.318}* \\
CODER(RB, RB)                & 0.634          & 0.316             \\ \bottomrule
\end{tabular}%
%}
}
% \end{adjustwidth}
\caption{Performance when applying CODER to RepBERT on the TripClick HEAD dataset, using multi-level (DCTR) relevance labels (metrics cut-off of 10). All CODER results are statistically significant (paired $t$-test, $p<0.05$) with respect to both the base and BM25$\rightarrow$base methods. The symbol * on best results denotes statistically significant improvement with respect to all baselines. Results with $^1$ are from \citet{rekabsaz_tripclick_2021}, with $^2$ from \citet{hofstatter_establishing_2022}.}\vspace{-10pt}
\label{tab:tripclick}
\end{table}

\vspace{-0.2cm}
\subsection{Results on TripClick}\label{sec:tripclick_results}

Following~\citet{rekabsaz_tripclick_2021}, we report performance in terms of MRR@10, nDCG@10 and Recall@10; nDCG@10 is considered the most important metric, as multiple relevant documents per query exist, and the DCTR relevance set additionally uses multiple levels of relevance. The results are presented in Table~\ref{tab:tripclick}: Using the same hyperparameters as in MS MARCO, CODER fine-tuning tremendously improves the performance of RepBERT trained on TripClick, both in reranking as well as in single-stage dense retrieval. The improvement is especially pronounced on the HEAD subset, where many relevance judgements per query are available. CODER achieves the SOTA performance by a large margin, ahead of all published results in the literature and on the TripClick leaderboard\footnote{https://tripdatabase.github.io/tripclick/}, including ensembles of heavyweight cross-encoder BERT Rerankers which were pre-trained on the domain-specific PubMedBERT (medicine) and SciBERT (science) corpora~\cite{hofstatter_establishing_2022}. CODER also significantly outperforms the best existing dense retrieval method (BERT-Dot pre-trained on SciBERT~\cite{hofstatter_establishing_2022}) on the TORSO and TAIL subsets (see Appendix Table~\ref{tab:tripclick_large}). However, presumably because these subsets consist of rare queries with significantly fewer relevance judgements per query, it falls behind the domain-pretrained cross-encoders.

Finally, we use TripClick's validation and test sets purely for ``zero-shot'' evaluating the RepBERT, CODER(RepBERT), CoCondenser and CODER(CoCondenser) models trained exclusively on MS MARCO. We emphasize that, uniquely in this setting, these models have not been trained or fine-tuned on TripClick; they are the same models described in Section~\ref{sec:msmarco}, now evaluated on a large dataset with severe distribution shift (biomedical domain).  While zero-shot evaluation is used increasingly often to demonstrate the effectiveness and generalizability of large transformer models (e.g. \cite{brown_language_2020, hao_language_2022, radford_learning_2021}), here we use it as a means to bypass the challenge of the expensive pre-training and fine-tuning of CoCondenser on TripClick. Results are shown in Appendix Table~\ref{tab:tripclick_zeroshot}: naturally, we observe that zero-shot performance in absolute terms is low; however, CODER-trained models perform always better. Moreover, the performance order CODER(CoCondenser) > CoCondenser > CODER(RepBERT) > RepBERT, consistent with our observations for MS MARCO, holds also here in almost every comparison.

\vspace{-0.2cm}
\subsection{Efficiency}
What is the additional cost of using CODER? Using a single NVIDIA TITAN RTX GPU (on a node with Intel Xeon Gold 6142 CPU), it takes about 186 ms to rank 1000 candidates per query in a batch of 32 queries (out of which less than 10ms refer to computing representations and scores, with the rest taken up by batching and loading samples to the GPU), i.e., a latency of 5.5-5.8 ms per query is introduced when using CODER as a second-stage reranker in a cascade.
Using CODER as a single-stage dense retriever only requires the same processing time as the base method, e.g. RepBERT or TAS-B. Table~\ref{tab:CODER_MSMARCO} reports the latency reported in \cite{zhan_repbert_2020}, but this in practice will be determined by the time for loading the query sequence to the GPU and encoding it (in our setup, approx. 5.4 ms per query,  out of which approx.\ 0.3 ms is the time to compute the representation), in addition to the time for finding the approximate nearest neighbors using a library such as FAISS~\cite{johnson_billion-scale_2017}. 

\subsection{Analysis of key factors}\label{sec:ablation}

\subsubsection{Importance of context}\label{sec:ablation_context}

In this section we wish to assess the intuitions that scoring a document within a context of other documents related to the same query can be advantageous, and that a list-wise loss function like the one we present in Section~\ref{sec:loss_function} is better equipped to leverage this context compared to a superposition of separate pair-wise loss components. We first study the impact of varying the type and number of negative documents during training.

\begin{figure}[t]
    \centering
    \includegraphics[width=1\linewidth]{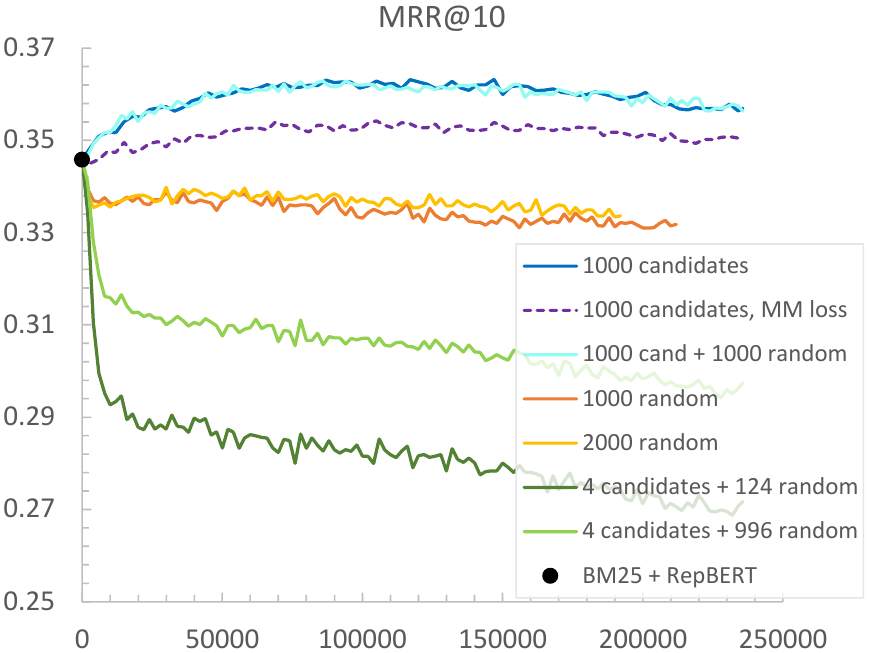}\\
    \includegraphics[width=1\linewidth]{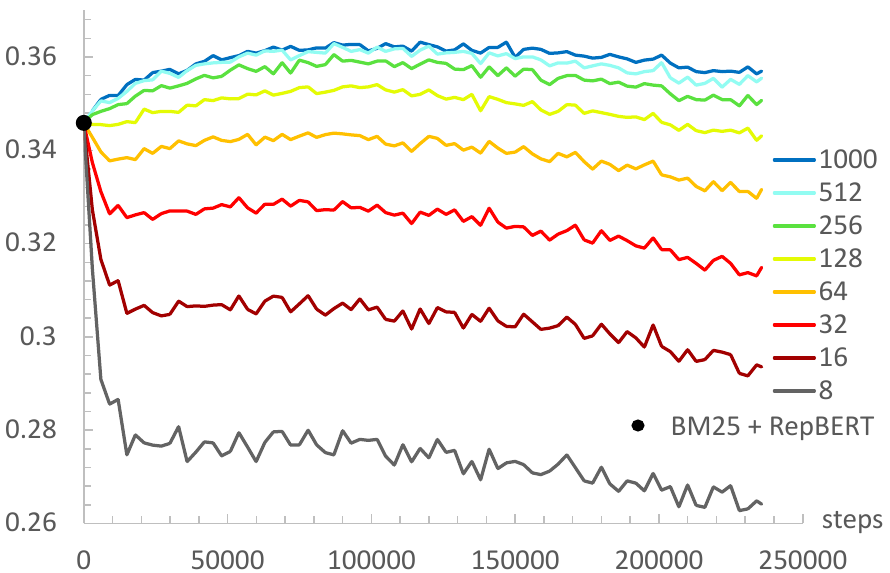}
    \caption{Performance of BM25$\rightarrow$CODER(RepBERT) on MS MARCO validation set over training steps. The left-most point corresponds to reranking BM25 candidates using the fully trained RepBERT. \textbf{Top}: Effect of type and number of documents used as negatives. The purple dashed curve corresponds to training with a pair-wise (Max Margin) loss.
    \textbf{Bottom}: Effect of number of BM25 candidates used as negatives during training.
    }\vspace{-10pt}
    \label{fig:CODER_RepBERT_MRR_vs_steps_composition_negatives_loss}
\end{figure}

In Figure~\ref{fig:CODER_RepBERT_MRR_vs_steps_composition_negatives_loss} we show how the performance of a model initialized from a trained RepBERT base model evolves during fine-tuning through CODER, measured in MRR@10 on our MS MARCO validation set when reranking 1000 candidates first retrieved for each query by BM25. Different curves correspond to different training settings.
% \footnote{Values tend to be higher than the ones in Table~\ref{tab:CODER_MSMARCO} due to evaluating on the validation set, and because the ground truth document is within the set of candidates to be reranked, even if BM25 failed to retrieve it.}
 The leftmost evaluation point corresponds to the best model checkpoint achievable through standard triplet-based training. 

\textbf{Composition of negatives:} We observe that training with only randomly sampled negatives, even in large numbers, leads to a deterioration of performance. We note that this is the case because through CODER we are optimizing a baseline model which has already been trained to peak performance through random negative documents. Using a small number of retrieved candidates together with in-batch random negatives, as in current SOTA methods~\cite{zhan_learning_2020, qu_rocketqa_2021}, again leads to deteriorating performance, even with many random negatives. Only by using a large number of coherent, query-specific negative samples (i.e. retrieved by a retrieval method) for the same query can CODER extract additional performance from the baseline. Moreover, adding 1000 additional random documents per query (in-batch negatives) on top of those does not yield any benefit, presumably because they are not adding any context (as they are unrelated to the query and documents under assessment) and/or they are not sufficiently challenging. 

\textbf{List-wise loss:} When training with the most commonly used pair-wise loss (Multi-Label Max Margin, purple dashed curve), even when using 1000 retrieved candidates as negatives within exactly the same training setup, i.e. including the same positive and negative examples for the same query in the same batch, we observe that performance only modestly improves performance (+0.009 MRR@10).
The improvement can be attributed to the fact that the model now encounters a large number of coherent, query-specific documents for the same query during a single step of training, which offers a more complete and accurate view of the loss landscape and thus leads to a more accurate update of parameters.
However, the ranking context is exploited more effectively when using a list-wise KL-divergence loss (dark blue curve), as CODER fine-tuning improves MRR@10 from 0.345 to 0.363 (+0.018).

Finally, we note that the training loss continuously decreased throughout training in all settings mentioned above (see Figure~\ref{fig:CODER_RepBERT_TrainLoss_vs_steps_composition_negatives_loss} in the Appendix), including the ones in which performance on the validation set was deteriorating at the same time (Figure~\ref{fig:CODER_RepBERT_MRR_vs_steps_composition_negatives_loss}). This fact suggests that deteriorating performance in those settings can be interpreted as overfitting, when there is insufficient information/signal captured by the training objective in order to learn to rank more effectively.

\subsubsection{Number of context documents}

We now investigate the importance of the quantity of negative documents per query during training. In Figure~\ref{fig:CODER_RepBERT_MRR_vs_steps_composition_negatives_loss}, where the number of BM25-retrieved candidate negatives is adjusted, we observe increasing performance per additional document until we reach a number in the order of the dimensionality of the embedding space (here, 768 for BERT-base). Increasing the number of negatives beyond this point yields diminishing returns (see section \ref{sec:large_number_negatives} for an explanation). 

We note that these numbers of negative \textit{candidates} (as opposed to random documents) per query are much higher than the ones used in all contemporary work (max.\ 4 candidates have been employed in \citet{qu_rocketqa_2021} and 30 in \citet{Gao2021UnsupervisedCA}); the reason that such a high number is necessary in order to achieve a performance improvement is that we are fine-tuning a model already trained to saturation. Only a large number of retrieved candidate negatives provides enough training signal to overcome overfitting and improve performance (Figure \ref{fig:CODER_RepBERT_TrainLoss_vs_steps_num_negatives} in the Appendix shows the training loss  to decrease in all cases).

%that when increasing the number of negatives, performance initially increases at a fast rate (i.e. performance per additional negative document), until we reach a number in the order of the dimensionality of the embedding space (here, 768 of BERT base), beyond which point we observe diminishing returns. To obtain a significant performance benefit over the base method, it is evident that a large number of negative documents (in the order of the dimensionality of space) is required, but this benefit clearly saturates as we keep adding negatives. We note that these numbers of negative \textit{candidates} (as opposed to random documents) per query are much higher than the ones used in all contemporary work (max. 4 candidates have been employed, in \citet{qu_rocketqa_2021}); the reason that such a high number is necessary in order to achieve a performance improvement is that we are fine-tuning a model already trained to saturation. Again, it appears that only a large number of candidate negatives provides enough training signal to overcome overfitting and improve performance, evidenced by the fact that in the rest of the settings, training loss was decreasing at the same time that performance on the validation was deteriorating (see Figure~\ref{fig:CODER_RepBERT_TrainLoss_vs_steps_num_negatives}).

The above results suggest that most dense retrieval models would likely benefit from training using a context, i.e., a large number of retrieved candidate documents per query, combined with an appropriate list-wise loss.

% \begin{figure}[h]
%     \centering
%     \includegraphics[width=1\linewidth]{CODER_RepBERT_MRR_vs_steps_composition_negatives_loss_CROPPED.png}
%     \caption{Performance of BM25$\rightarrow$CODER(RepBERT, BM25) (i.e. when reranking 1000 BM25 candidates) on the validation set, as a function of training steps. The left-most point corresponds to reranking BM25 candidates using the fully trained RepBERT. Different curves correspond to a different number of BM25 candidates used as negatives during training. 
%     %Only using a large number of retrieved candidates as negatives results in a significant improvement of performance over the base method.
%     }
%     \label{fig:CODER_RepBERT_MRR_vs_steps_num_negatives}
% \end{figure}

\section{Conclusion}

We examine the importance of ranking context and the effect of its constituent parts, i.e., a fully list-wise loss, a large number of negatives, and retrieved (query-specific) instead of random negatives, and show that they are all important ingredients for improving performance. We demonstrate that a lightweight reranking framework designed to leverage context is sufficient to significantly enhance the effectiveness of a wide range of dense retrieval models, without expensive cross-encoder distillation, pseudo-labeling or ``denoising'' negatives. The only computational overhead is a fast, resource-light fine-tuning process, with little (when the model is used as a reranker) to no (when used as a single-stage retriever) extra computational cost during inference.
%This framework is premised on training expensive transformer-based retrieval models through the efficient use of a large number of cheap, query-specific negative samples per query within a list-wise loss function. 

% We show that this approach achieves substantial improvement over competitive base methods and over scoring candidate documents in isolation from one another, while offering computational efficiency.
%Our results suggest that most dense retrieval models would likely benefit from training using such a reranking context.
%, i.e., a large number of retrieved (rather than random) candidate documents per query, combined with an appropriate list-wise loss.
%Future work should address explicitly capturing inter-document relationships, which cannot be trivially modeled through a stack of self-attention layers.% as well account for list-wise properties such as fairness and bias.
% Finally, scoring multiple candidate documents simultaneously for the same query enables additional possibilities with respect to the calibration of relevance estimates, as well as mitigating societal bias in ranking.

\section{Limitations}

 In the present work we have endeavored to demonstrate that the performance benefits achieved by our framework are generalizable with respect to the dense retrieval method used as base, as well as the document collections to be searched. For this purpose, we selected as baselines dual encoder models which are highly representative of the state-of-the-art IR approaches, including ones that have undergone highly optimized training processes.
 %Also, besides the commonly used MS MARCO and TREC Deep Learning track datasets (the latter usually being too small to establish statistical significance), we have trained and evaluated our models on TripClick, which is a healthcare IR dataset with distinct characteristics: domain-specific terminology, queries predominantly by expert users (healthcare professionals), numerous relevance judgement points per document and multiple levels of relevance.

However, it is always possible that some retrieval models will not enjoy similar performance enhancement when trained through CODER, owing to the large variety of modeling assumptions or specially required conditions for training different models. For example, a model trained through a process already leveraging most of the ranking context elements examined here (list-wise loss, query-specific/retrieved candidates, a large number of them) will stand to benefit less. Conversely, models trained on datasets including multiple positive documents per query will exhibit better performance.

We further emphasize that the retrieval methods compatible with our approach make use of \textit{distinct} document and query representations. Thus, it is not possible to use CODER to directly fine-tune term-interaction (a.k.a. cross-encoder) models like BERT Reranker, because each document within CODER is only available as a single embedding vector. Finally, we note that, although it is theoretically possible to initialize the query encoder from a generic NLP language model, in all our presented experiments it is initialized from a base model (query/document encoder) already trained for retrieval, the one that is also used to precompute document embeddings.

\section*{Acknowledgements}
G. Zerveas would like to thank the Onassis Foundation for supporting this research. This work received financial support from the State of Upper Austria and the Federal Ministry of Education, Science, and Research, through grant LIT-2021-YOU-215, and is also supported in part by the NSF (IIS-1956221). The views and conclusions contained herein are those of the authors and should not be interpreted as necessarily representing the official policies, either expressed or implied, of NSF or the U.S.\ Government.

\FloatBarrier

\bibliography{reference}

\begin{thebibliography}{55}
\expandafter\ifx\csname natexlab\endcsname\relax\def\natexlab#1{#1}\fi

\bibitem[{Ai et~al.(2018)Ai, Bi, Guo, and Croft}]{ai_learning_2018}
Qingyao Ai, Keping Bi, Jiafeng Guo, and W.~Bruce Croft. 2018.
\newblock \href {https://doi.org/10.1145/3209978.3209985} {Learning a {Deep}
  {Listwise} {Context} {Model} for {Ranking} {Refinement}}.
\newblock In \emph{The 41st {International} {ACM} {SIGIR} {Conference} on
  {Research} \& {Development} in {Information} {Retrieval}}, {SIGIR} '18, pages
  135--144, New York, NY, USA. Association for Computing Machinery.

\bibitem[{Ai et~al.(2019)Ai, Wang, Bruch, Golbandi, Bendersky, and
  Najork}]{ai_learning_2019}
Qingyao Ai, Xuanhui Wang, Sebastian Bruch, Nadav Golbandi, Michael Bendersky,
  and Marc Najork. 2019.
\newblock \href {http://arxiv.org/abs/1811.04415} {Learning {Groupwise}
  {Multivariate} {Scoring} {Functions} {Using} {Deep} {Neural} {Networks}}.
\newblock \emph{arXiv:1811.04415 [cs]}.
\newblock ArXiv: 1811.04415.

\bibitem[{Bajaj et~al.(2018)Bajaj, Campos, Craswell, Deng, Gao, Liu, Majumder,
  McNamara, Mitra, Nguyen, Rosenberg, Song, Stoica, Tiwary, and
  Wang}]{bajaj_ms_2018}
Payal Bajaj, Daniel Campos, Nick Craswell, Li~Deng, Jianfeng Gao, Xiaodong Liu,
  Rangan Majumder, Andrew McNamara, Bhaskar Mitra, Tri Nguyen, Mir Rosenberg,
  Xia Song, Alina Stoica, Saurabh Tiwary, and Tong Wang. 2018.
\newblock \href {http://arxiv.org/abs/1611.09268} {{MS} {MARCO}: {A} {Human}
  {Generated} {MAchine} {Reading} {COmprehension} {Dataset}}.
\newblock \emph{arXiv:1611.09268 [cs]}.
\newblock ArXiv: 1611.09268.

\bibitem[{Brown et~al.(2020)Brown, Mann, Ryder, Subbiah, Kaplan, Dhariwal,
  Neelakantan, Shyam, Sastry, Askell, Agarwal, Herbert-Voss, Krueger, Henighan,
  Child, Ramesh, Ziegler, Wu, Winter, Hesse, Chen, Sigler, Litwin, Gray, Chess,
  Clark, Berner, McCandlish, Radford, Sutskever, and
  Amodei}]{brown_language_2020}
Tom~B. Brown, Benjamin Mann, Nick Ryder, Melanie Subbiah, Jared Kaplan,
  Prafulla Dhariwal, Arvind Neelakantan, Pranav Shyam, Girish Sastry, Amanda
  Askell, Sandhini Agarwal, Ariel Herbert-Voss, Gretchen Krueger, Tom Henighan,
  Rewon Child, Aditya Ramesh, Daniel~M. Ziegler, Jeffrey Wu, Clemens Winter,
  Christopher Hesse, Mark Chen, Eric Sigler, Mateusz Litwin, Scott Gray,
  Benjamin Chess, Jack Clark, Christopher Berner, Sam McCandlish, Alec Radford,
  Ilya Sutskever, and Dario Amodei. 2020.
\newblock \href {https://doi.org/10.48550/arXiv.2005.14165} {Language {Models}
  are {Few}-{Shot} {Learners}}.
\newblock ArXiv:2005.14165 [cs].

\bibitem[{Bruch et~al.(2020)Bruch, Han, Bendersky, and
  Najork}]{bruch_stochastic_l2r2020}
Sebastian Bruch, Shuguang Han, Mike Bendersky, and Marc Najork. 2020.
\newblock A stochastic treatment of learning to rank scoring functions.
\newblock In \emph{Proceedings of the 13th ACM International Conference on Web
  Search and Data Mining (WSDM 2020)}, pages 61--69.

\bibitem[{Burges(2010)}]{Burges2010FromRT}
Christopher J.~C. Burges. 2010.
\newblock From ranknet to lambdarank to lambdamart: An overview.

\bibitem[{Calauz\`{e}nes et~al.(2012)Calauz\`{e}nes, Usunier, and
  Gallinari}]{ranking_loss_calibration_neurips2012}
Cl\'{e}ment Calauz\`{e}nes, Nicolas Usunier, and Patrick Gallinari. 2012.
\newblock \href
  {https://proceedings.neurips.cc/paper/2012/file/50f3f8c42b998a48057e9d33f4144b8b-Paper.pdf}
  {On the (non-)existence of convex, calibrated surrogate losses for ranking}.
\newblock In \emph{Advances in Neural Information Processing Systems},
  volume~25. Curran Associates, Inc.

\bibitem[{Cao et~al.(2007)Cao, Qin, Liu, Tsai, and Li}]{cao_learning_2007}
Zhe Cao, Tao Qin, Tie-Yan Liu, Ming-Feng Tsai, and Hang Li. 2007.
\newblock \href {https://doi.org/10.1145/1273496.1273513} {Learning to rank:
  from pairwise approach to listwise approach}.
\newblock In \emph{Proceedings of the 24th international conference on
  {Machine} learning}, {ICML} '07, pages 129--136, New York, NY, USA.
  Association for Computing Machinery.

\bibitem[{Chen et~al.(2022)Chen, Hui, He, Han, Sun, and
  Ye}]{chen_incorporating_2022}
Xiaoyang Chen, Kai Hui, Ben He, Xianpei Han, Le~Sun, and Zheng Ye. 2022.
\newblock \href {https://doi.org/10.1007/978-3-030-99736-6_8} {Incorporating
  {Ranking} {Context} for {End}-to-{End} {BERT} {Re}-ranking}.
\newblock In \emph{Advances in {Information} {Retrieval}}, Lecture {Notes} in
  {Computer} {Science}, pages 111--127, Cham. Springer International
  Publishing.

\bibitem[{Chen and Eickhoff(2021)}]{Chen2021PoolRankMP}
Zhizhong Chen and Carsten Eickhoff. 2021.
\newblock Poolrank: Max/min pooling-based ranking loss for listwise learning \&
  ranking balance.
\newblock \emph{ArXiv}, abs/2108.03586.

\bibitem[{Craswell et~al.(2021)Craswell, Mitra, Yilmaz, and
  Campos}]{craswell2021overview}
Nick Craswell, Bhaskar Mitra, Emine Yilmaz, and Daniel Campos. 2021.
\newblock Overview of the trec 2020 deep learning track.
\newblock \emph{arXiv preprint arXiv:2102.07662}.

\bibitem[{Craswell et~al.(2020)Craswell, Mitra, Yilmaz, Campos, and
  Voorhees}]{craswell2020overview}
Nick Craswell, Bhaskar Mitra, Emine Yilmaz, Daniel Campos, and Ellen~M
  Voorhees. 2020.
\newblock Overview of the {TREC} 2019 deep learning track.
\newblock \emph{arXiv preprint arXiv:2003.07820}.

\bibitem[{Devlin et~al.(2019)Devlin, Chang, Lee, and
  Toutanova}]{devlin_bert_2019}
Jacob Devlin, Ming-Wei Chang, Kenton Lee, and Kristina Toutanova. 2019.
\newblock \href {https://doi.org/10.18653/v1/N19-1423} {{BERT}: {Pre}-training
  of {Deep} {Bidirectional} {Transformers} for {Language} {Understanding}}.
\newblock In \emph{{NAACL}}.

\bibitem[{Gao and Callan(2021{\natexlab{a}})}]{Gao2021CondenserAP}
Luyu Gao and Jamie Callan. 2021{\natexlab{a}}.
\newblock Condenser: a pre-training architecture for dense retrieval.
\newblock In \emph{EMNLP}.

\bibitem[{Gao and Callan(2021{\natexlab{b}})}]{Gao2021UnsupervisedCA}
Luyu Gao and Jamie Callan. 2021{\natexlab{b}}.
\newblock Unsupervised corpus aware language model pre-training for dense
  passage retrieval.
\newblock \emph{ArXiv}, abs/2108.05540.

\bibitem[{Gao et~al.(2021)Gao, Dai, and Callan}]{gao2021coil}
Luyu Gao, Zhuyun Dai, and Jamie Callan. 2021.
\newblock Coil: Revisit exact lexical match in information retrieval with
  contextualized inverted list.
\newblock In \emph{Conference of the North American Chapter of the Association
  for Computational Linguistics: Human Language Technologies}.

\bibitem[{Hao et~al.(2022)Hao, Song, Dong, Huang, Chi, Wang, Ma, and
  Wei}]{hao_language_2022}
Yaru Hao, Haoyu Song, Li~Dong, Shaohan Huang, Zewen Chi, Wenhui Wang, Shuming
  Ma, and Furu Wei. 2022.
\newblock \href {https://doi.org/10.48550/arXiv.2206.06336} {Language {Models}
  are {General}-{Purpose} {Interfaces}}.
\newblock ArXiv:2206.06336 [cs].

\bibitem[{Hofstätter et~al.(2022)Hofstätter, Althammer, Sertkan, and
  Hanbury}]{hofstatter_establishing_2022}
Sebastian Hofstätter, Sophia Althammer, Mete Sertkan, and Allan Hanbury. 2022.
\newblock \href {http://arxiv.org/abs/2201.00365} {Establishing {Strong}
  {Baselines} for {TripClick} {Health} {Retrieval}}.
\newblock \emph{arXiv:2201.00365 [cs]}.
\newblock ArXiv: 2201.00365.

\bibitem[{Hofstätter et~al.(2021)Hofstätter, Lin, Yang, Lin, and
  Hanbury}]{hofstatter_efficiently_2021}
Sebastian Hofstätter, Sheng-Chieh Lin, Jheng-Hong Yang, Jimmy~J. Lin, and
  A.~Hanbury. 2021.
\newblock \href {https://doi.org/10.1145/3404835.3462891} {Efficiently
  {Teaching} an {Effective} {Dense} {Retriever} with {Balanced} {Topic} {Aware}
  {Sampling}}.
\newblock \emph{SIGIR}.

\bibitem[{Joachims(2002)}]{joachims_optimizing_2002}
T.~Joachims. 2002.
\newblock \href
  {https://www.semanticscholar.org/paper/Optimizing-search-engines-using-clickthrough-data-Joachims/cfd4259d305a00f13d5f08841230389f61322422}
  {Optimizing search engines using clickthrough data}.
\newblock \emph{undefined}.

\bibitem[{Johnson et~al.(2017)Johnson, Douze, and
  Jégou}]{johnson_billion-scale_2017}
Jeff Johnson, Matthijs Douze, and Hervé Jégou. 2017.
\newblock \href {http://arxiv.org/abs/1702.08734} {Billion-scale similarity
  search with {GPUs}}.
\newblock \emph{arXiv:1702.08734 [cs]}.
\newblock ArXiv: 1702.08734.

\bibitem[{Karpukhin et~al.(2020)Karpukhin, Oguz, Min, Lewis, Wu, Edunov, Chen,
  and Yih}]{karpukhin_dense_2020}
Vladimir Karpukhin, Barlas Oguz, Sewon Min, Patrick Lewis, Ledell Wu, Sergey
  Edunov, Danqi Chen, and Wen-tau Yih. 2020.
\newblock \href {https://doi.org/10.18653/v1/2020.emnlp-main.550} {Dense
  {Passage} {Retrieval} for {Open}-{Domain} {Question} {Answering}}.
\newblock In \emph{Proceedings of the 2020 {Conference} on {Empirical}
  {Methods} in {Natural} {Language} {Processing} ({EMNLP})}, pages 6769--6781,
  Online. Association for Computational Linguistics.

\bibitem[{Khattab and Zaharia(2020)}]{khattab_colbert_2020}
Omar Khattab and Matei Zaharia. 2020.
\newblock {ColBERT}: {Efficient} and {Effective} {Passage} {Search} via
  {Contextualized} {Late} {Interaction} over {BERT}.
\newblock In \emph{Proceedings of the 43rd International ACM SIGIR conference
  on research and development in Information Retrieval}, pages 39--48.

\bibitem[{Lesota et~al.(2021)Lesota, Rekabsaz, Cohen, Grasserbauer, Eickhoff,
  and Schedl}]{10.1145/3471158.3472229}
Oleg Lesota, Navid Rekabsaz, Daniel Cohen, Klaus~Antonius Grasserbauer, Carsten
  Eickhoff, and Markus Schedl. 2021.
\newblock \href {https://doi.org/10.1145/3471158.3472229} {\emph{A Modern
  Perspective on Query Likelihood with Deep Generative Retrieval Models}}, page
  185–195. Association for Computing Machinery, New York, NY, USA.

\bibitem[{Lin et~al.(2021)Lin, Nogueira, and Yates}]{lin2021pretrained}
Jimmy Lin, Rodrigo Nogueira, and Andrew Yates. 2021.
\newblock Pretrained transformers for text ranking: Bert and beyond.
\newblock \emph{Synthesis Lectures on Human Language Technologies},
  14(4):1--325.

\bibitem[{Lin et~al.(2020)Lin, Yang, and Lin}]{lin_distilling_2020}
Sheng-Chieh Lin, Jheng-Hong Yang, and Jimmy Lin. 2020.
\newblock \href {http://arxiv.org/abs/2010.11386} {Distilling {Dense}
  {Representations} for {Ranking} using {Tightly}-{Coupled} {Teachers}}.
\newblock \emph{arXiv:2010.11386 [cs]}.
\newblock ArXiv: 2010.11386.

\bibitem[{Liu et~al.(2019)Liu, Ott, Goyal, Du, Joshi, Chen, Levy, Lewis,
  Zettlemoyer, and Stoyanov}]{liu_roberta_2019}
Yinhan Liu, Myle Ott, Naman Goyal, Jingfei Du, Mandar Joshi, Danqi Chen, Omer
  Levy, Mike Lewis, Luke Zettlemoyer, and Veselin Stoyanov. 2019.
\newblock \href {http://arxiv.org/abs/1907.11692} {{RoBERTa}: {A} {Robustly}
  {Optimized} {BERT} {Pretraining} {Approach}}.
\newblock \emph{arXiv:1907.11692 [cs]}.
\newblock ArXiv: 1907.11692.

\bibitem[{Luan et~al.(2021)Luan, Eisenstein, Toutanova, and
  Collins}]{luan_sparse_2021}
Yi~Luan, Jacob Eisenstein, Kristina Toutanova, and Michael Collins. 2021.
\newblock \href {https://doi.org/10.1162/tacl_a_00369} {Sparse, {Dense}, and
  {Attentional} {Representations} for {Text} {Retrieval}}.
\newblock \emph{Transactions of the Association for Computational Linguistics},
  9:329--345.

\bibitem[{Mallia et~al.(2021)Mallia, Khattab, Suel, and
  Tonellotto}]{mallia2021learning}
Antonio Mallia, Omar Khattab, Torsten Suel, and Nicola Tonellotto. 2021.
\newblock Learning passage impacts for inverted indexes.
\newblock In \emph{Proceedings of the 44th International ACM SIGIR Conference
  on Research and Development in Information Retrieval}, pages 1723--1727.

\bibitem[{Mitra and Craswell(2018)}]{mitra_introduction_2018}
Bhaskar Mitra and Nick Craswell. 2018.
\newblock \href {https://doi.org/10.1561/1500000061} {An {Introduction} to
  {Neural} {Information} {Retrieval} t}.
\newblock \emph{Foundations and Trends® in Information Retrieval},
  13(1):1--126.

\bibitem[{Naseri et~al.(2021)Naseri, Dalton, Yates, and Allan}]{naseri2021ceqe}
Shahrzad Naseri, Jeffrey Dalton, Andrew Yates, and James Allan. 2021.
\newblock Ceqe: Contextualized embeddings for query expansion.
\newblock In \emph{European Conference on Information Retrieval}, pages
  467--482. Springer.

\bibitem[{Nogueira and Cho(2020)}]{nogueira_passage_2020}
Rodrigo Nogueira and Kyunghyun Cho. 2020.
\newblock \href {http://arxiv.org/abs/1901.04085} {Passage {Re}-ranking with
  {BERT}}.
\newblock \emph{arXiv:1901.04085 [cs]}.
\newblock ArXiv: 1901.04085.

\bibitem[{Nogueira et~al.(2020)Nogueira, Jiang, Pradeep, and
  Lin}]{nogueira2020document}
Rodrigo Nogueira, Zhiying Jiang, Ronak Pradeep, and Jimmy Lin. 2020.
\newblock Document ranking with a pretrained sequence-to-sequence model.
\newblock In \emph{Findings of the Association for Computational Linguistics:
  EMNLP 2020}, pages 708--718.

\bibitem[{Pang et~al.(2020)Pang, Xu, Ai, Lan, Cheng, and
  Wen}]{pang_setrank_2020}
Liang Pang, Jun Xu, Qingyao Ai, Yanyan Lan, Xueqi Cheng, and Jirong Wen. 2020.
\newblock \href {https://doi.org/10.1145/3397271.3401104} {{SetRank}:
  {Learning} a {Permutation}-{Invariant} {Ranking} {Model} for {Information}
  {Retrieval}}.
\newblock In \emph{Proceedings of the 43rd {International} {ACM} {SIGIR}
  {Conference} on {Research} and {Development} in {Information} {Retrieval}},
  {SIGIR} '20, pages 499--508, New York, NY, USA. Association for Computing
  Machinery.

\bibitem[{Pasumarthi et~al.(2019{\natexlab{a}})Pasumarthi, Bruch, Wang, Li,
  Bendersky, Najork, Pfeifer, Golbandi, Anil, and
  Wolf}]{pasumarthi_tf-ranking_2019}
Rama~Kumar Pasumarthi, Sebastian Bruch, Xuanhui Wang, Cheng Li, Michael
  Bendersky, Marc Najork, Jan Pfeifer, Nadav Golbandi, Rohan Anil, and Stephan
  Wolf. 2019{\natexlab{a}}.
\newblock \href {https://doi.org/10.1145/3292500.3330677} {{TF}-{Ranking}:
  {Scalable} {TensorFlow} {Library} for {Learning}-to-{Rank}}.
\newblock In \emph{Proceedings of the 25th {ACM} {SIGKDD} {International}
  {Conference} on {Knowledge} {Discovery} \& {Data} {Mining}}, {KDD} '19, pages
  2970--2978, New York, NY, USA. Association for Computing Machinery.

\bibitem[{Pasumarthi et~al.(2019{\natexlab{b}})Pasumarthi, Wang, Bendersky, and
  Najork}]{pasumarthi_self-attentive_2019}
Rama~Kumar Pasumarthi, Xuanhui Wang, Michael Bendersky, and Marc Najork.
  2019{\natexlab{b}}.
\newblock \href {http://arxiv.org/abs/1910.09676} {Self-{Attentive} {Document}
  {Interaction} {Networks} for {Permutation} {Equivariant} {Ranking}}.
\newblock \emph{arXiv:1910.09676 [cs]}.
\newblock ArXiv: 1910.09676.

\bibitem[{Qu et~al.(2021)Qu, Ding, Liu, Liu, Ren, Zhao, Dong, Wu, and
  Wang}]{qu_rocketqa_2021}
Yingqi Qu, Yuchen Ding, Jing Liu, Kai Liu, Ruiyang Ren, Wayne~Xin Zhao, Daxiang
  Dong, Hua Wu, and Haifeng Wang. 2021.
\newblock \href {http://arxiv.org/abs/2010.08191} {{RocketQA}: {An} {Optimized}
  {Training} {Approach} to {Dense} {Passage} {Retrieval} for {Open}-{Domain}
  {Question} {Answering}}.
\newblock \emph{arXiv:2010.08191 [cs]}.
\newblock ArXiv: 2010.08191.

\bibitem[{Radford et~al.(2021)Radford, Kim, Hallacy, Ramesh, Goh, Agarwal,
  Sastry, Askell, Mishkin, Clark, Krueger, and
  Sutskever}]{radford_learning_2021}
Alec Radford, Jong~Wook Kim, Chris Hallacy, Aditya Ramesh, Gabriel Goh,
  Sandhini Agarwal, Girish Sastry, Amanda Askell, Pamela Mishkin, Jack Clark,
  Gretchen Krueger, and Ilya Sutskever. 2021.
\newblock \href {https://doi.org/10.48550/arXiv.2103.00020} {Learning
  {Transferable} {Visual} {Models} {From} {Natural} {Language} {Supervision}}.
\newblock ArXiv:2103.00020 [cs].

\bibitem[{Rekabsaz et~al.(2021{\natexlab{a}})Rekabsaz, Kopeinik, and
  Schedl}]{rekabsaz2021societal}
Navid Rekabsaz, Simone Kopeinik, and Markus Schedl. 2021{\natexlab{a}}.
\newblock Societal biases in retrieved contents: Measurement framework and
  adversarial mitigation for bert rankers.
\newblock In \emph{Proceedings of the 44th International ACM SIGIR Conference
  on Research and Development in Information Retrieval}.

\bibitem[{Rekabsaz et~al.(2021{\natexlab{b}})Rekabsaz, Lesota, Schedl, Brassey,
  and Eickhoff}]{rekabsaz_tripclick_2021}
Navid Rekabsaz, Oleg Lesota, Markus Schedl, Jon Brassey, and Carsten Eickhoff.
  2021{\natexlab{b}}.
\newblock \href {https://doi.org/10.1145/3404835.3463242} {{TripClick}: {The}
  {Log} {Files} of a {Large} {Health} {Web} {Search} {Engine}}.
\newblock In \emph{Proceedings of the 44th {International} {ACM} {SIGIR}
  {Conference} on {Research} and {Development} in {Information} {Retrieval}},
  pages 2507--2513. Association for Computing Machinery, New York, NY, USA.

\bibitem[{Rekabsaz and Schedl(2020)}]{rekabsaz2020neural}
Navid Rekabsaz and Markus Schedl. 2020.
\newblock Do neural ranking models intensify gender bias?
\newblock In \emph{Proceedings of the 43rd International ACM SIGIR Conference
  on Research and Development in Information Retrieval}, pages 2065--2068.

\bibitem[{Ren et~al.(2021{\natexlab{a}})Ren, Lv, Qu, Liu, Zhao, She, Wu, Wang,
  and Wen}]{ren_pair_2021}
Ruiyang Ren, Shangwen Lv, Yingqi Qu, Jing Liu, Wayne~Xin Zhao, QiaoQiao She,
  Hua Wu, Haifeng Wang, and Ji-Rong Wen. 2021{\natexlab{a}}.
\newblock \href {https://doi.org/10.18653/v1/2021.findings-acl.191} {{PAIR}:
  {Leveraging} {Passage}-{Centric} {Similarity} {Relation} for {Improving}
  {Dense} {Passage} {Retrieval}}.
\newblock \emph{Findings of the Association for Computational Linguistics:
  ACL-IJCNLP 2021}, pages 2173--2183.
\newblock ArXiv: 2108.06027.

\bibitem[{Ren et~al.(2021{\natexlab{b}})Ren, Qu, Liu, Zhao, She, Wu, Wang, and
  Wen}]{ren_rocketqav2_2021}
Ruiyang Ren, Yingqi Qu, Jing Liu, Wayne~Xin Zhao, QiaoQiao She, Hua Wu, Haifeng
  Wang, and Ji-Rong Wen. 2021{\natexlab{b}}.
\newblock \href {https://doi.org/10.18653/v1/2021.emnlp-main.224}
  {{RocketQAv2}: {A} {Joint} {Training} {Method} for {Dense} {Passage}
  {Retrieval} and {Passage} {Re}-ranking}.
\newblock In \emph{Proceedings of the 2021 {Conference} on {Empirical}
  {Methods} in {Natural} {Language} {Processing}}, pages 2825--2835, Online and
  Punta Cana, Dominican Republic. Association for Computational Linguistics.

\bibitem[{Sanh et~al.(2020)Sanh, Debut, Chaumond, and
  Wolf}]{sanh_distilbert_2020}
Victor Sanh, Lysandre Debut, Julien Chaumond, and Thomas Wolf. 2020.
\newblock \href {http://arxiv.org/abs/1910.01108} {{DistilBERT}, a distilled
  version of {BERT}: smaller, faster, cheaper and lighter}.
\newblock \emph{arXiv:1910.01108 [cs]}.
\newblock ArXiv: 1910.01108.

\bibitem[{Santhanam et~al.(2021)Santhanam, Khattab, Saad-Falcon, Potts, and
  Zaharia}]{santhanam_colbertv2_2021}
Keshav Santhanam, Omar Khattab, Jon Saad-Falcon, Christopher Potts, and Matei
  Zaharia. 2021.
\newblock \href {https://doi.org/10.48550/arXiv.2112.01488} {{ColBERTv2}:
  {Effective} and {Efficient} {Retrieval} via {Lightweight} {Late}
  {Interaction}}.
\newblock ArXiv:2112.01488 [cs].

\bibitem[{Vaswani et~al.(2017)Vaswani, Shazeer, Parmar, Uszkoreit, Jones,
  Gomez, Kaiser, and Polosukhin}]{vaswani_attention_2017}
Ashish Vaswani, Noam Shazeer, Niki Parmar, Jakob Uszkoreit, Llion Jones,
  Aidan~N Gomez, Łukasz Kaiser, and Illia Polosukhin. 2017.
\newblock Attention is {All} you {Need}.
\newblock In I.~Guyon, U.~V. Luxburg, S.~Bengio, H.~Wallach, R.~Fergus,
  S.~Vishwanathan, and R.~Garnett, editors, \emph{Advances in {Neural}
  {Information} {Processing} {Systems} 30}, pages 5998--6008. Curran
  Associates, Inc.

\bibitem[{Xiong et~al.(2020)Xiong, Xiong, Li, Tang, Liu, Bennett, Ahmed, and
  Overwijk}]{xiong_approximate_2020}
Lee Xiong, Chenyan Xiong, Ye~Li, Kwok-Fung Tang, Jialin Liu, Paul Bennett,
  Junaid Ahmed, and Arnold Overwijk. 2020.
\newblock \href {http://arxiv.org/abs/2007.00808} {Approximate {Nearest}
  {Neighbor} {Negative} {Contrastive} {Learning} for {Dense} {Text}
  {Retrieval}}.
\newblock \emph{arXiv:2007.00808 [cs]}.
\newblock ArXiv: 2007.00808.

\bibitem[{Yang et~al.(2017)Yang, Fang, and Lin}]{yang_anserini_2017}
Peilin Yang, Hui Fang, and Jimmy Lin. 2017.
\newblock \href {https://doi.org/10.1145/3077136.3080721} {Anserini: {Enabling}
  the {Use} of {Lucene} for {Information} {Retrieval} {Research}}.
\newblock In \emph{Proceedings of the 40th {International} {ACM} {SIGIR}
  {Conference} on {Research} and {Development} in {Information} {Retrieval}},
  {SIGIR} '17, pages 1253--1256, New York, NY, USA. Association for Computing
  Machinery.

\bibitem[{Yanjun~Ma and Yanjun~Ma(2019)}]{yanjun_ma_paddlepaddle_2019}
Dianhai~Yu Yanjun~Ma and Dianhai~Yu Yanjun~Ma. 2019.
\newblock \href {https://doi.org/10.11871/jfdc.issn.2096.742X.2019.01.011}
  {{PaddlePaddle}: {An} {Open}-{Source} {Deep} {Learning} {Platform} from
  {Industrial} {Practice}}.
\newblock \emph{Frontiers of Data and Domputing}, 1(1):105--115.

\bibitem[{Zhan et~al.(2021{\natexlab{a}})Zhan, Mao, Liu, Guo, Zhang, and
  Ma}]{zhan2021jointly}
Jingtao Zhan, Jiaxin Mao, Yiqun Liu, Jiafeng Guo, Min Zhang, and Shaoping Ma.
  2021{\natexlab{a}}.
\newblock \href {https://doi.org/10.1145/3459637.3482358} {\emph{Jointly
  Optimizing Query Encoder and Product Quantization to Improve Retrieval
  Performance}}, page 2487–2496. Association for Computing Machinery, New
  York, NY, USA.

\bibitem[{Zhan et~al.(2021{\natexlab{b}})Zhan, Mao, Liu, Guo, Zhang, and
  Ma}]{zhan_optimizing_2021}
Jingtao Zhan, Jiaxin Mao, Yiqun Liu, Jiafeng Guo, Min Zhang, and Shaoping Ma.
  2021{\natexlab{b}}.
\newblock \href {https://doi.org/10.1145/3404835.3462880} {Optimizing {Dense}
  {Retrieval} {Model} {Training} with {Hard} {Negatives}}.
\newblock In \emph{Proceedings of the 44th {International} {ACM} {SIGIR}
  {Conference} on {Research} and {Development} in {Information} {Retrieval}},
  pages 1503--1512, Virtual Event Canada. ACM.

\bibitem[{Zhan et~al.(2020{\natexlab{a}})Zhan, Mao, Liu, Zhang, and
  Ma}]{zhan_learning_2020}
Jingtao Zhan, Jiaxin Mao, Yiqun Liu, Min Zhang, and Shaoping Ma.
  2020{\natexlab{a}}.
\newblock \href {http://arxiv.org/abs/2010.10469} {Learning {To} {Retrieve}:
  {How} to {Train} a {Dense} {Retrieval} {Model} {Effectively} and
  {Efficiently}}.
\newblock \emph{arXiv:2010.10469 [cs]}.
\newblock ArXiv: 2010.10469.

\bibitem[{Zhan et~al.(2020{\natexlab{b}})Zhan, Mao, Liu, Zhang, and
  Ma}]{zhan_repbert_2020}
Jingtao Zhan, Jiaxin Mao, Yiqun Liu, Min Zhang, and Shaoping Ma.
  2020{\natexlab{b}}.
\newblock \href {http://arxiv.org/abs/2006.15498} {{RepBERT}: {Contextualized}
  {Text} {Embeddings} for {First}-{Stage} {Retrieval}}.
\newblock \emph{arXiv:2006.15498 [cs]}.
\newblock ArXiv: 2006.15498.

\bibitem[{Zhang et~al.(2019)Zhang, Han, Liu, Jiang, Sun, and
  Liu}]{zhang_ernie_2019}
Zhengyan Zhang, Xu~Han, Zhiyuan Liu, Xin Jiang, Maosong Sun, and Qun Liu. 2019.
\newblock \href {https://doi.org/10.18653/v1/P19-1139} {{ERNIE}: {Enhanced}
  {Language} {Representation} with {Informative} {Entities}}.
\newblock In \emph{Proceedings of the 57th {Annual} {Meeting} of the
  {Association} for {Computational} {Linguistics}}, pages 1441--1451, Florence,
  Italy. Association for Computational Linguistics.

\bibitem[{Zheng et~al.(2020)Zheng, Hui, He, Han, Sun, and
  Yates}]{zheng2020bert}
Zhi Zheng, Kai Hui, Ben He, Xianpei Han, Le~Sun, and Andrew Yates. 2020.
\newblock {BERT-QE}: Contextualized query expansion for document re-ranking.
\newblock In \emph{Findings of the Association for Computational Linguistics:
  EMNLP 2020}, pages 4718--4728.

\end{thebibliography}
\bibliographystyle{acl_natbib}

\appendix
\section{Appendix}
\label{sec:appendix}

\subsection{Motivation for simultaneously scoring a large number of negative documents}\label{sec:large_number_negatives}

\begin{figure}[h]
    \centering
    \includegraphics[width=1\linewidth]{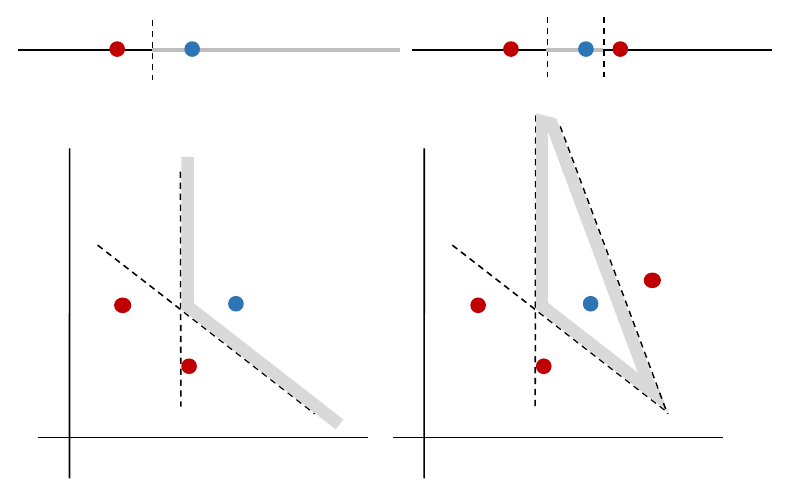}
    \caption{Positive (blue) and negative (red) document vectors in a shared document and query embedding space of 1 (upper row) and 2 (lower row) dimensions. The contrastive learning objective simply requires the query representation to be mapped closer to the positive than all negative documents; this means that for a $d$-dimensional space, the query representation need not necessarily lie in proximity to the positive document, but simply within an infinite subspace (left column: grey part of line in 1D space, grey-outlined part of plane in 2D space). At least $d+1$ negative documents are required to constrain the space of favorable query mappings to a bounded convex polytope. Given that the positive document is contained within a simplex formed by $d+1$ negative documents as vertices in that space (interval in $\mathbb{R}$, triangle in $\mathbb{R}^2$, tetrahedron in $\mathbb{R}^3$ etc), the loss function will favor mapping the query onto another simplex in the same space (right column: grey interval in 1D space, grey-outlined triangular area in 2D space).}
    \label{fig:negatives_and_spaces}
\end{figure}

 In contrastive learning, models are trained to assign a higher score for similarity between the query and a positive document than between the query and all negative documents: $s(q, p_j^+) > s(q, p_i^-),~\forall i,j$, where $ p_j^+$ denotes a positive and $ p_i^-$ a negative document/passage respectively. In Figure~\ref{fig:negatives_and_spaces}, vector representations of documents are depicted as points (red for positive, blue for negative documents) in a $d$-dimensional space that is shared with the query representation vector. 
 Because functions used to compute similarity increase with decreasing Euclidean distance, the objective can be fulfilled by learning to map the query within a smaller distance from a given positive document $ p_j^+$ compared to all negative documents\footnote{Because document vectors are distributed far from the origin, this will typically be true even when the dot product is used as a similarity function.}. However, for a space of dimension $d$, when fewer than $d+1$ negative documents are included in a loss calculation, there is an infinite subspace where the query representation can lie, arbitrarily far from the positive document, and still satisfy this condition. At least $d+1$ negative documents are required to constrain the space of objective-favored query mappings to a bounded convex polytope: given that the positive document is contained within a simplex formed by $d+1$ negative documents as vertices in that space (interval in $\mathbb{R}$, triangle in $\mathbb{R}^2$, tetrahedron in $\mathbb{R}^3$ etc), the loss function will favor mapping the query onto another simplex in the same space, within which the distance from the positive document will be bounded. Of course, training with fewer than $d+1$ negative documents per query still works: although the loss landscape as revealed by only a few negative documents is a rough approximation, and thus the parameter updates computed by stochastic gradient descent for each batch will be suboptimal and noisy, a good minimum of the loss can still be found in expectation by iterating over the entire training set. However, as the number of negative documents per query increases to $d+1$, the approximations of the gradients of the loss and thus parameter updates at each training step will be more accurate and therefore training will be more efficient. Increasing the number of negatives to an even higher number is still expected to yield a performance improvement, but at a reduced rate: this is because (a) there is no guarantee that with $d+1$ negatives per query, the positive document will be contained within a simplex of negatives, but with every additional negative this probability increases, and (b) once the positive document is contained within a finite simplex, the probability for every additional negative document to further constraint the bounded subspace where the loss can be minimized becomes increasingly smaller.

The above theoretical analysis can explain the observations made e.g. by \citet{hofstatter_efficiently_2021}, who obtained significantly better performance when increasing the batch size from 32 to 256, or by ~\citet{qu_rocketqa_2021}, who used ``cross-batch'' random negatives (pooled from different GPUs) to effectively increase the number of ``in-batch'' negatives to a few thousand documents in order to ``reduce the discrepancy between training and inference", and noticed a substantial improvement of performance as a result. Also, in agreement to our explanation above, the performance improvement they observed as a function of the number of negatives quickly saturated when exceeding the dimensionality of the document embedding space. 

Finally, it is evident from the analysis above that the closer the negative document representations lie to to the ground truth representation (i.e. the more relevant the negatives are), the smaller the bounded convex subspace will be, a fact which supports the observed importance of challenging negatives in literature~\cite{xiong_approximate_2020, zhan_optimizing_2021, qu_rocketqa_2021, hofstatter_efficiently_2021}.

\subsection{Data}\label{sec:data_appendix}

In the present work we only examine passage retrieval, and use the terms ``passage'' and ``document'' interchangeably. All data used are in English.

\subsubsection{MS MARCO and TREC Deep Learning}
Following the standard practice in related contemporary literature, we use the MS MARCO dataset~\cite{bajaj_ms_2018}, which has been sourced from open-domain logs of the Bing search engine, for training and evaluating our models. The MS MARCO passage collection contains about 8.8 million documents and the training set contains about 503k queries labeled with one or (rarely) more relevant documents, on a single level of relevance.

For validation we use a subset of 10k samples from ``MS MARCO dev'', which is a set containing about 56k labeled queries, and refer to it as ``MS MARCO dev 10k''. As a test set we use a different, officially designated subset of ``MS MARCO dev'', originally called ``MS MARCO dev.small'', which contains 6980 queries. However, following standard practice in literature and leaderboards, we refer to it as ``MS MARCO dev''. We also evaluate on the TREC Deep Learning track 2019 and 2020 test sets, each containing 43 and 54 queries respectively, labeled to an average ``depth'' of more than 210 document judgements per query, and using 4 levels of relevance: ``Not Relevant'' (0), ``Related'' (1), ``Highly Relevant'' (2) and ``Perfect'' (3). According to the official (strict) interpretation of relevance labels\footnote{\url{https://trec.nist.gov/data/deep2019.html}}, a level of 1 should not be considered relevant and thus be treated just like a level of 0, while the lenient interpretation considers passages of level 1 relevant  when calculating metrics.

\subsubsection{TripClick}

We additionally evaluate our framework on a different dataset, for two reasons: first, to assess the robustness and generality of our framework across datasets. Second, our approach is premised on using a list-wise loss function while jointly scoring a large number of documents within the same query context, and thus differences from pair-wise approaches are expected to be more pronounced when training on a dataset where several documents have been judged with respect to their relevance to a query.

TripClick is a recently introduced health IR dataset~\cite{rekabsaz_tripclick_2021} based on click logs that refer to about 1.5M MEDLINE articles. The approx. 700k unique queries in its training set are split into 3 subsets, HEAD, TORSO and TAIL, based on their frequency of occurrence: queries in TAIL are asked only once or a couple of times, while queries in HEAD have been asked tens or hundreds of times. As a result, each query in HEAD, TORSO and TAIL on average ends up with 41.9, 9.1 and 2.8 pseudo-relevance judgements, using a click-through model (RAW) where every clicked document is considered relevant. The dataset also includes alternative relevance judgements using the Document Click-Through Rate (DCTR), on 4 distinct levels (the latter follow the same definitions as the TREC Deep Learning evaluation sets). For validation and evaluation of our models we use the officially designated validation and test set, respectively (3.5k queries each).

\begin{table}
\center
\begin{tabular}{@{}ll@{}}
\toprule
\textbf{Parameter}     & \textbf{Value} \\ \midrule
Max. query length      & 32             \\
Max. doc. length (MS MARCO)       & 256            \\
Max. doc. length (TripClick)       & 512            \\
Batch size             & 32             \\
Optimizer              & RAdam          \\
Adam epsilon           & 1.3e-7         \\
Learning rate          & 1.73e-6        \\
LR warmup steps        & 9000           \\
Weight decay           & 9.5e-5         \\
Dropout                & 0.1            \\
Max. gradient clipping & 1.0            \\
$d_{\text{model}}$     & 768            \\ \bottomrule
\end{tabular}
\caption{Main configuration parameters of CODER (without transformation of document representations).}
\label{tab:coder_config}
\end{table}

\begin{figure*}[h]
    \centering
    \includegraphics[width=1\linewidth]{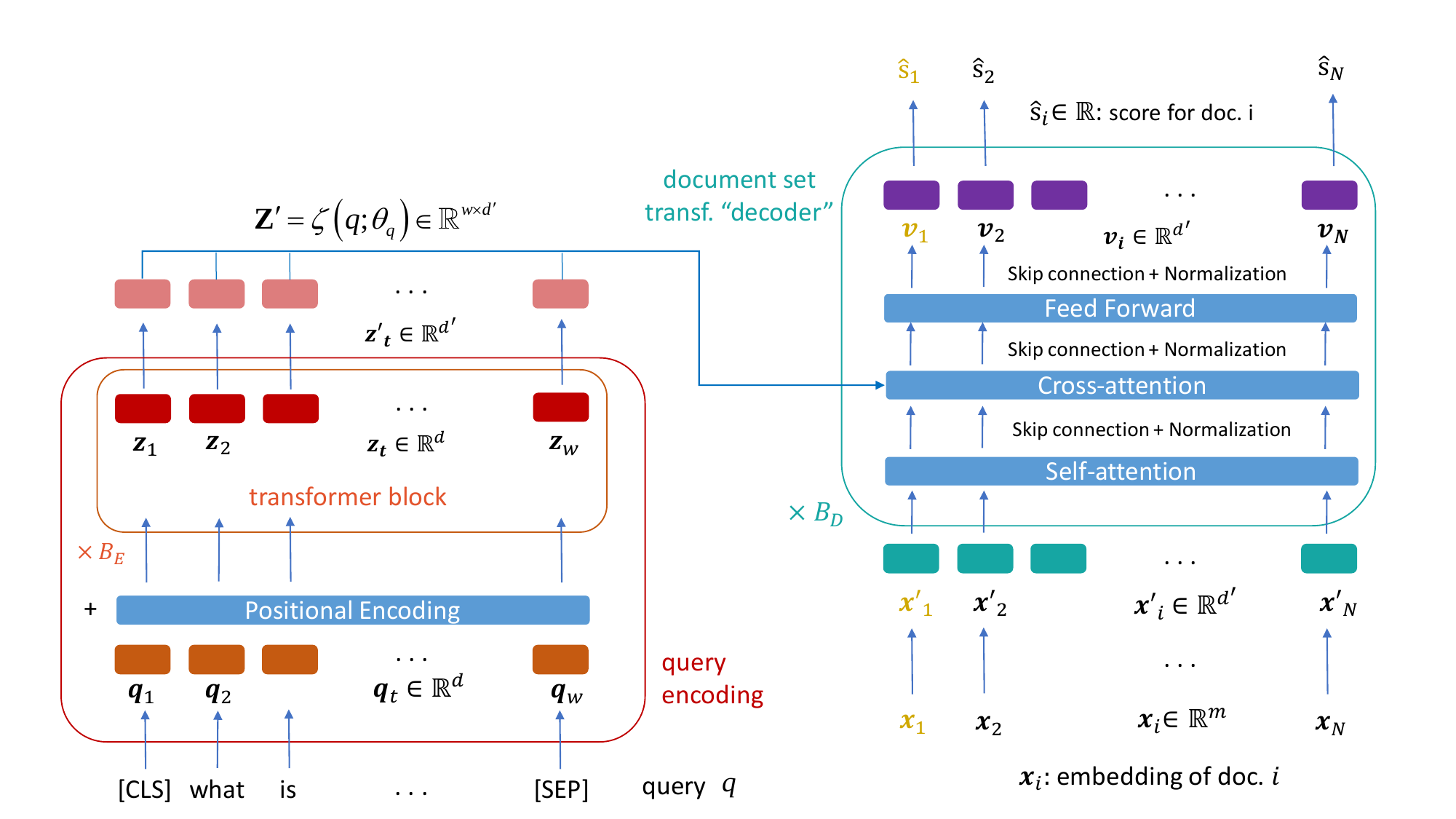}
    \caption{Schematic diagram of CODER employing a transformer ``decoder'' as a document set scoring module.}
    \label{fig:tranf_coder_diagram}
\end{figure*}

\begin{figure}[h]
    \centering
    \includegraphics[width=1\linewidth]{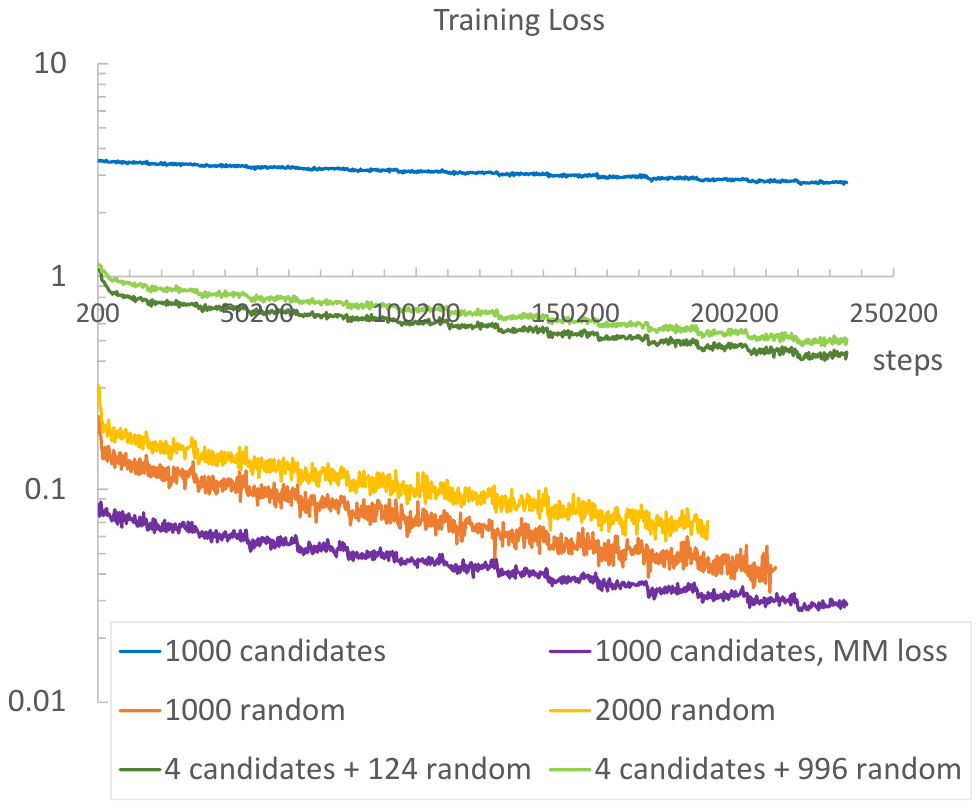}
    \caption{Evolution of the loss on the training set, as training of BM25$\rightarrow$CODER(RepBERT) progresses. Different curves correspond to a different type and number of documents used as negatives during training. While training loss is decreasing in all cases, only using numerous retrieved candidates as negatives, combined with a list-wise KL-divergence loss, results in a significant improvement of performance over the base method (see Figure~\ref{fig:CODER_RepBERT_MRR_vs_steps_composition_negatives_loss}) on the validation and test sets.}
    \label{fig:CODER_RepBERT_TrainLoss_vs_steps_composition_negatives_loss}
\end{figure}

\begin{figure}[h]
    \centering
    \includegraphics[width=1\linewidth]{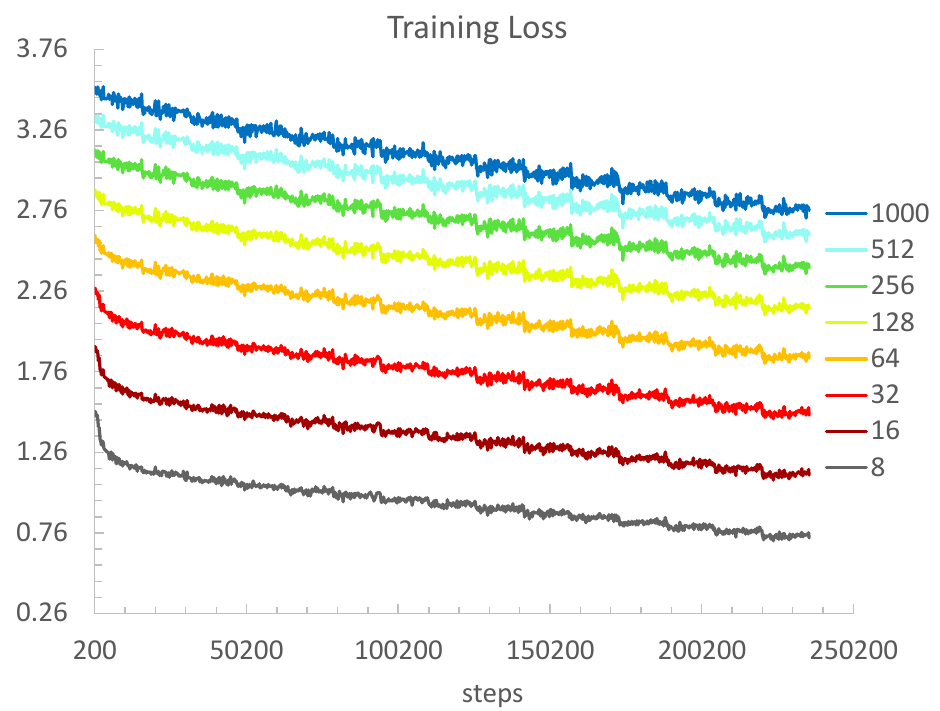}
    \caption{Evolution of the loss on the training set, as training of BM25$\rightarrow$CODER(RepBERT) progresses. Different curves correspond to a different number of BM25 candidates used as negatives during training. While training loss is decreasing in all cases, only using a large number of candidates results in a significant improvement of performance over the base method on the validation and test sets (see Figure~\ref{fig:CODER_RepBERT_MRR_vs_steps_composition_negatives_loss}).}
    \label{fig:CODER_RepBERT_TrainLoss_vs_steps_num_negatives}
\end{figure}

\subsection{Evaluation}\label{sec:evaluation}

We use mean reciprocal rank (MRR), normalized discounted cumulative gain (nDCG), and recall to evaluate the models on TREC DL tracks, MS MARCO and TripClick, in line with past work (e.g.~\cite{xiong_approximate_2020, zhan_learning_2020, hofstatter_efficiently_2021, rekabsaz_tripclick_2021}). All training and evaluation experiments are produced with the same seed for pseudo-random number generators. While relevance judgements are well-defined in MS MARCO and TripClick, for the TREC DL tracks there exist strict and lenient interpretations of the relevance scores of judged documents (see Section~\ref{sec:data_appendix}). As past work has been inconsistent in specifying which interpretation is used, we evaluate all models on both versions, denoted as \{lenient\}/\{strict\} in Table~\ref{tab:CODER_MSMARCO}. We calculate the metrics using the official TREC evaluation software.\footnote{\url{trec.nist.gov/trec_eval/index.html}}

\subsection{Using a transformer as document scoring function}\label{sec:transformer_document_scoring}

We have experimented with more complex, parametric functions $\varphi$ as scoring modules, including feedforward neural networks and stacks of transformer blocks (see diagram in Figure~\ref{fig:tranf_coder_diagram}), which explicitly model inter-document relationships. When we use a transformer encoder as a scoring module, we do not use positional encodings over the input document embeddings, because we require permutation invariance: given fixed positional encodings of ranking order, it would be trivial for the model to learn to assign higher scores for documents of higher rank; at the same time, there is no meaningful sequence order over document embeddings other than relevance ranking. 

Although a transformer encoder of 2 blocks was able to marginally outperform the simple function of Equation~\eqref{eq:predicted_scores}, the small improvement makes it hard to justify precluding the option of single-stage dense retrieval, as well as the additional computational cost. However, in future work, we intend to more thoroughly investigate the use of a transformer-based scoring function, adding more transformer blocks, which we have not been able to implement due to computational constraints - self-attention incurs a $O(N^2)$ GPU memory dependence on the number of context documents $N$.

\subsection{How context can help even without parametric modeling of document relationships}\label{sec:context_without_doc_relationships}

 The improvement of our contextual reranking framework over standard triplet training can be attributed to the fact that the model now encounters many more query-specific candidate documents for the same query during a single step of training, which offers a more complete and precise view of the loss landscape and thus leads to a more accurate update of parameters in the same step (also see discussion in Sec.~\ref{sec:large_number_negatives}). However, this constitutes a “weak” exploitation of ranking context offered by CODER. 
 
 The ranking context is exploited more effectively when using a list-wise KL divergence loss. We believe that one reason is that often, among the retrieved candidates, some documents which have not been labeled as positive, are in fact relevant (i.e. false/mislabeled negatives)~\cite{qu_rocketqa_2021}. The KL-divergence function is well-positioned to deal with this case, compared to a pair-wise loss (e.g. Max-margin loss)  and the Negative Loglikelihood Loss (NLL, a.k.a. InfoNCE) as used e.g. in \cite{karpukhin_dense_2020} and \cite{qu_rocketqa_2021}. The KL-divergence loss, which compares the distribution of predicted scores against the annotated relevance scores, does not directly penalize assigning a high score to a document annotated as non-relevant; instead, it severely penalizes assigning a low score to a ground-truth relevant document. Therefore, as long as the ground-truth positive document $p$ receives a not-too-low normalized relevance score $\hat{s}_p=\text{softmax}(\varphi(\mathbf{X}))_p$, e.g. $\hat{s}_p>0.2$, which allows it to escape the very steep part of the loss curve $L(\hat{s}_p) = -\log (\hat{s}_p)$ close to $\hat{s}_p=0$, the loss will still be small. Thus, it will not severely affect the model’s parameters to erroneously force ranking $p$ higher than the false negatives.
 
 Of course, the more such false negative documents exist, receiving non-zero weight in the predicted score distribution, the more difficult it becomes for $\hat{s}_p$ to be high, which leads to a higher loss. However, the existence of more than one labeled positives $(k>1)$ per query alleviates this problem: the individual normalized scores of the $k$ positives will be affected less by the existence of false negatives, and because of the form of $L(x) = -\log (x)$, the overall loss will be smaller than in the case of $k=1$. This is an additional benefit the KL-divergence loss has over NLL, as the NLL only takes into account a single positive document at a time.
% For reference, we show below the difference between KL-divergence (left) and NLL (right side of the inequality):

% \begin{equation*}
% \begin{split}
% {\cal L}\left( {{\bf{s}},{\bf{\hat s}}} \right) &= {{\rm{D}}_{{\rm{KL}}}}\left( {\sigma ({\bf{y}})||\sigma ({\bf{\hat s}})} \right) \\
% &=  - \sum\limits_{j = 1}^{N = k + n} {\sigma {{({\bf{y}})}_j}} \log \frac{{\sigma {{({\bf{\hat s}})}_j}}}{{\sigma {{({\bf{y}})}_j}}} \\
% &=  - \sum\limits_{j = 1}^k {\frac{1}{k}} \log \frac{{\user2{\sigma }\left( {{\rm{sim}}({q_i},p_j^ + )} \right)}}{{{\raise0.5ex\hbox{$\scriptstyle 1$}
% \kern-0.1em/\kern-0.15em
% \lower0.25ex\hbox{$\scriptstyle k$}}}} \\
% &= - \frac{1}{k}\sum\limits_{j = 1}^k {\log k\frac{{\exp \left( {{\rm{sim}}({q_i},p_j^ + )} \right)}}{{\sum\limits_{l= 1}^k {\exp \left( {{\rm{sim}}({q_i},p_l^ + )} \right) + } \sum\limits_{l = 1}^n {\exp \left( {{\rm{sim}}({q_i},p_l^ - )} \right)} }}}  \\
% &\ne  - \left[ {\frac{1}{{{B_s}}}} \right]\log \frac{{\exp \left( {{\rm{sim}}({q_i},p_j^ + )} \right)}}{{\exp \left( {{\rm{sim}}({q_i},p_l^ + )} \right) + \sum\limits_{l = 1}^n {\exp \left( {{\rm{sim}}({q_i},p_l^ - )} \right)} }}    
% \end{split}
% \end{equation*}

\subsection{Notes on RepBERT used for reranking BM25}\label{sec:repbert_vs_bm25}

We surprisingly found RepBERT to be significantly more effective when used as a reranker with BM25 as the first stage retrieval method, although it has been developed and so far only considered and evaluated as a stand-alone dense retrieval method. This is despite RepBERT being an overall much stronger retrieval method than BM25, with a much better recall (0.943 with a cut-off at 1000 passages on MS MARCO dev, versus 0.853 for BM25). This observation may fall within the scope of the general discussion regarding the advantages of combining exact (lexical) and inexact (latent representation) matching~\cite{mitra_introduction_2018}. One can also hypothesize that BM25 may filter out candidates which would otherwise (spuriously or justifiably) lie close to the ground truth passsage in a latent semantic space, and would have been thus preferred by the deep learning ranker, while not being considered as ``hits'' in the metrics; this interpretation is supported by the observations of \cite{qu_rocketqa_2021}, who found that very often actually relevant passages have not been labeled as such in MS MARCO.

\begin{table*}[t]
\centering
\resizebox{1\textwidth}{!}{%
{\setlength\doublerulesep{0.5pt}
\begin{tabular}{@{}l|ll|lll|lll|lll@{}}
\toprule
%\multicolumn{1}{c|}{\textbf{TripClick TEST}} 
\multirow{2}{*}{\textbf{Model}} 
& \multicolumn{2}{c|}{\textbf{DCTR Head}}                & \multicolumn{3}{c|}{\textbf{RAW Head}}               & \multicolumn{3}{c|}{\textbf{RAW Torso}}              & \multicolumn{3}{c}{\textbf{RAW Tail}}                \\ 
& \textbf{MRR}   & \textbf{nDCG} & \textbf{MRR} & \textbf{nDCG} & \textbf{Recall} & \textbf{MRR} & \textbf{nDCG} & \textbf{Recall} & \textbf{MRR} & \textbf{nDCG} & \textbf{Recall} \\ \midrule

BM25$^1$                     & 0.276         & 0.224   & - &	0.199	& 0.128		& - & 0.206	& 0.262	& - &	0.267	& 0.409    \\
Transformer-Kernel$^1$                  & -         & 0.284 & -	& 0.284	& 0.167	& -	& 0.272	& 0.321	& -	& 0.295	 & 0.459                  \\ 
BERT-Dot (SciBERT)$^2$                  & 0.530          & 0.243 & - &	- 	& -	& - &	-	&-	&-	&-	& -         \\
BERT-Cat (SciBERT)$^2$                  & 0.595   & 0.294 & - &	- & -	& \textbf{0.459}	& \textbf{0.360}	& -	& \textbf{0.377}	& \textbf{0.408}	& -  \\
\midrule \midrule %\midrule
RepBERT  (abbrev: RB)                       & 0.526          & 0.255   & 0.574 &	0.344 &	0.199 &	0.338 &	0.246 &	0.309	& 0.254 &	0.268 &	0.404       \\
BM25 $\rightarrow$ RepBERT                      & 0.538        & 0.262    & 0.592 &	0.356 &	0.204 &	0.359 &	0.269	& 0.340 &	0.278 &	0.297 &	0.445      \\ \cdashlinelr{1-12}%\midrule
RB $\rightarrow$ CODER(RB, RB)         & \textbf{0.637}* & \textbf{0.318}*  & \textbf{0.679}* &	\textbf{0.421}* &	\textbf{0.235}* &	0.433 &	0.308 &	0.355 &	0.296 &	0.315 &	0.469 \\
CODER(RB, RB)                & 0.634          & 0.316     & 0.674 &	0.419 &	0.234 &	0.433 &	0.308 &	0.355 & 0.296 &	0.315 &	0.468       \\ \bottomrule

\end{tabular}%
}
}
% \end{adjustwidth}
\caption{Performance when applying CODER to RepBERT on the TripClick dataset, using multi-level (DCTR) and binary (RAW) relevance labels (cut-off of 10). The symbol * on best results denotes statistically significant (paired $t$-test, $p<0.05$) improvement with respect to all baselines. Results with $^1$ are from \citet{rekabsaz_tripclick_2021}, with $^2$ from \citet{hofstatter_establishing_2022}.}
\label{tab:tripclick_large}
\end{table*}

\begin{table*}[t]
\centering
\resizebox{1.0\textwidth}{!}{%
%{\setlength\doublerulesep{0.5pt}
%\small
\begin{tabular}{@{}l|ll|lll|lll|lll@{}}
\toprule
%\multicolumn{1}{c|}{\textbf{TripClick TEST}} 
\multirow{2}{*}{\textbf{Model}} 
& \multicolumn{2}{c|}{\textbf{DCTR Head}}                & \multicolumn{3}{c|}{\textbf{RAW Head}}               & \multicolumn{3}{c|}{\textbf{RAW Torso}}              & \multicolumn{3}{c}{\textbf{RAW Tail}}                \\ 
& \textbf{MRR}   & \textbf{nDCG} & \textbf{MRR} & \textbf{nDCG} & \textbf{Recall} & \textbf{MRR} & \textbf{nDCG} & \textbf{Recall} & \textbf{MRR} & \textbf{nDCG} & \textbf{Recall} \\ \midrule

RepBERT-MSM                      &    0.233     &  0.107  & 0.278 &	0.149 &	0.085 & 	0.205 &	0.130 &	0.151 &	0.117 &	0.122 &	0.195  \\
CODER-MSM(RepBERT-MSM)          & 0.244 & 0.113 & 0.294 &	0.157 &	0.091 &	0.211 &	0.139 &	0.166 &	0.127 &	0.137 &	0.223 \\
% \cdashlinelr{1-12}%\midrule
Cocondenser-MSM  & 0.242          & 0.114       & 0.293 &	0.156 &	0.091 &	\textbf{0.217} &	0.144 &	0.178 &	0.153 &	0.162 &	0.254   \\
CODER-MSM(Cocondenser-MSM)                & \textbf{0.251}          & \textbf{0.117}  & \textbf{0.305} &	\textbf{0.161} &	\textbf{0.093}	& 0.216 &	\textbf{0.146} &	\textbf{0.182} &	\textbf{0.154} &	\textbf{0.164} &	\textbf{0.259}           \\ \bottomrule
\end{tabular}%
%}
}
% \end{adjustwidth}
\caption{Zero-shot results: all models are trained on MS MARCO (`-MSM' suffix) but evaluated on TripClick.}
\label{tab:tripclick_zeroshot}
\end{table*}

\subsection{Additional Results on TripClick}
Table~\ref{tab:tripclick_large} shows the performance of models trained and evaluated on the TripClick dataset. Table~\ref{tab:tripclick_zeroshot} shows zero-shot evaluation results for models trained on MS MARCO, but evaluated on TripClick (a dataset of queries and articles in the biomedical domain) without any additional training.

\subsection{Detailed comparison with related work}

Recent work on ad-hoc information retrieval has employed transformer-based architectures following two main approaches: the first approach~\cite{nogueira_passage_2020, khattab_colbert_2020} allows direct interactions between query and document terms through attention, offering impressive retrieval performance, albeit at the expense of computational efficiency; it can practically be used only as part of a cascade system for reranking candidate documents retrieved by a first-stage method, and still introduces a significant end-to-end processing delay, in the order of seconds per query. These powerful but slow ``cross-encoder'' models have also been used as teachers for ``dual encoder'' (a.k.a ``bi-encoder'') methods~\cite{lin_distilling_2020, qu_rocketqa_2021, ren_pair_2021, hofstatter_efficiently_2021}. These methods constitute the second approach, described below.

The dual encoder approach employs an architecture of two transformer encoders (optionally sharing weights, or implemented as the same encoder, distinguishing between queries and documents by adding special sequence type encodings) to separately encode the query and document sequences, without interactions between them. For inference, it relies on the efficient computation of the dot product through high-performing Approximate Nearest Neighbors libraries such as FAISS~\cite{johnson_billion-scale_2017} to evaluate the similarity between extracted query and document representations. This approach, called ``dense retrieval'', is highly effective, fast and single-stage, but still lags behind in terms of retrieval performance compared to the first approach.

Our framework addresses the current dilemma of slow reranking versus fast but less effective single-stage retrieval: it efficiently fine-tunes the query encoder of an existing (base) dense retrieval dual encoder model through reranking a set of precomputed document embeddings, offering a substantial performance improvement over the base model. During inference, it can be used either for rapid reranking in a cascade system, incurring a delay per query of a few milliseconds, or directly for single-stage dense retrieval, at the same speed as the base method, if no non-linear contextual document transformation is employed.

The state of the art has advanced through the exploration of techniques which select better negative documents so as to improve the training process. It has been clearly demonstrated that the quality of negative documents (essentially, how challenging/relevant they are) significantly affects similarity learning and confers performance benefits. Throughout training, \citet{xiong_approximate_2020} use the progressively improving document encoder to periodically (every few thousand steps) recompute document representations of all documents in the collection and re-index the document collection using FAISS. They also use the query encoder to recompute all training queries and retrieve the hardest (i.e. highest similarity) negatives through FAISS to construct triplets for the subsequent training period. \citet{zhan_learning_2020} improve on this effective but very slow and resource-intense process by eschewing fine-tuning of the document encoder, and they instead precompute and fix all document representations; they only fine-tune the query encoder, by retrieving at the end of each training step the hard negatives used for the next training step. Their motivation is two-fold: on the one hand, they argue that performing retrieval through the entire collection at each training step, instead of reranking candidates, reduces the discrepancy between inference retrieval and training tasks, naming their method ``Learning to Retrieve'' (published as ``STAR''/``ADORE'' in \cite{zhan_optimizing_2021}). On the other hand, they believe that reranking statically retrieved negatives quickly ceases to be effective, because the model quickly learns to rank static negatives lowly. Although we follow their approach in precomputing document embeddings and fine-tuning only the query encoder, in the present work we show that reranking statically retrieved negatives can indeed be a very effective framework, provided that one establishes a \textit{context} for the query: to achieve this, different from \citet{zhan_learning_2020}, who essentially use dynamically created triplets and a pair-wise loss, we use a large number ($N=1000$) of static pre-retrieved candidates per query (instead of a single or a couple) in combination with a list-wise loss function. Additionally, when using a parametric scoring function $\varphi$, our framework allows transforming document embeddings on the fly based on the context of the other candidate documents and the query itself.

The work of \citet{qu_rocketqa_2021} is also motivated by reducing the discrepancy between training and inference task, and as a solution they increase the quantity of negatives by using a computational framework for parallelism, PaddlePaddle~\cite{yanjun_ma_paddlepaddle_2019}, to multiply the number of in-batch negatives by the number of GPUs used for training (an approach they call ``cross-batch negatives''). As part of an elaborate, multi-step process, they fine-tune a base dual encoder model which has initially been trained through ``cross-batch'' negatives, after first retrieving the most relevant documents per query to use as static negatives. However, their approach only uses 4 such hard negatives per query, together with a pair-wise negative loglikelihood loss, and relies on several thousands of random cross-batch negatives. It thus differs from ours in that it does not establish a context; in fact, \citet{qu_rocketqa_2021} find that without training a slow but powerful term-interaction model for ``denoising'' (excluding the most relevant documents from the set of negatives), the performance of the fine-tuned model is substantially worse than the base model. Instead, our framework significantly improves performance when fine-tuning the base model, while being much faster and less demanding in terms of infrastructure. Also, since the number of negatives per query in our case is decoupled from the batch size, it can grow without the possibly detrimental effects that increasing the batch size can have on batch gradient descent optimization.

TAS-B~\cite{hofstatter_efficiently_2021} goes into the orthogonal direction of improving the quality of negative documents, using a simple but effective idea: it first clusters queries by semantic similarity and then packs queries from the same cluster into the same batch, such that their corresponding ground-truth relevant documents, which are used as in-batch negatives, are no longer random, as documents highly relevant to a query are very likely to also be relevant to the semantically related queries in the same batch. Despite starting from a different motivation,
%for a dataset such as MS MARCO (labeled with a single level of relevance and, most often, a single ground-truth relevant document per query), 
TAS-B can be seen as following an indirect, ``noisier'' version of our approach: effectively, the scores of several somewhat related in-batch negative documents enter the calculation of the loss for a given query. One difference lies in the type of loss function used: pair-wise in the case of TAS-B, list-wise in our case.
This becomes especially important in the case of more than one ground-truth relevant documents (rare in MS MARCO), and/or several levels of relevance judgements.
A second, more important difference is that our negatives are the top-$N-k$ documents retrieved by the base retrieval method (where $k$ is the number of positives), and thus for competitive base retrieval models they will almost certainly be of higher relevance than even the related in-batch negatives of TAS-B; at the same time, they are the documents which the model itself considers as the most similar to the query, and thus by definition the most challenging and suitable negatives to be juxtaposed with the ground-truth relevant document.
%Additionally, the CODER framework allows training for fairness, which is not possible directly through in-batch negatives, as well as the contextual transformation of each document representation before scoring.
%Finally, unlike CODER, TAS-B cannot be combined with ANCE/L2Re to benefit from dynamic retrieval of better negatives as the query representations improve (and document representations, in the case of ANCE), because it relies on using the ground-truth relevant documents of related/clustered queries as "hard negatives". Naturally, the ground-truth relevant documents don't change as as query representations improve, however the relevance of negative documents retrieved by ANCE/L2Re using the updated query representations can significantly improve throughout training. The only way TAS-B could benefit from the improving query encoder is by periodically recomputing all query representations in the dataset, indexing them and performing clustering anew, in a way similar to how ANCE is refreshing the document representations, but more computationally demanding because of clustering. Even in this case, any benefits would be indirect and discrete, since they could only manifest if the query representations would have changed enough to affect the assignment of queries into different clusters, which in turn would affect the composition of in-batch negatives in some batches.

With respect to simultaneously scoring a set of candidate documents, our approach is reminiscent of list-wise ranking models which employ self-attention~\cite{pang_setrank_2020, pasumarthi_self-attentive_2019}. However, there are several key differences:
\begin{itemize}
    \item Type of features: to evaluate semantic similarity between query and documents, these methods rely on BM25/TF-IDF interaction between query and document terms. These sparse query-document features, alongside PageRank and features such as freshness and click-through rates, are fed as input vectors to stack of customized self-attention blocks (SetRank,~\citep{pang_setrank_2020}) or to a single self-attention layer over document representation followed by concatenation with a query representation and scoring layers~\cite{pasumarthi_self-attentive_2019}.
    \item Architecture: our framework supports but does not necessarily rely on self-attention or other non-linear transformation of document representations. In the variant where we do make use of self-attention, we do not use positional encoding, in order to enforce permutation invariance. Not only do we represent queries very differently (through a separate query transformer encoder), but training the query encoder is the main objective of our method. Also, query representations can enter as part of a cross-attention module, which allows document representations to interact with query term representations.
    %In order to make SetRank compatible with MS MARCO, we would need to inject query representations into the input of the set transformer encoder.
    Thus, our architecture is akin to a complete encoder-decoder transformer~\cite{vaswani_attention_2017}: the encoder extracts query \textit{term} representations, while a ``decoder'' component (without attention masking or positional encoding) concurrently transforms and scores a set of \textit{document embeddings} (see Figure~\ref{fig:tranf_coder_diagram} in the Appendix).
    \item Annotation and learning objective: SetRank uses a custom ``attention rank loss function'' over provided target rankings, since the datasets used for training (Istella~LETOR\footnote{http://blog.istella.it/istella-learning-to-rank-dataset}, Microsoft~LETOR\footnote{http://research.microsoft.com/en-us/projects/mslr}, Yahoo!~LETOR\footnote{http://learningtorankchallenge.yahoo.com}) include 46-500 judged documents per query, on 5 levels of relevance.
    \item They are only used for reranking, with the pool of candidate documents given in the dataset. Our framework allows both reranking retrieved candidates of a first-stage retrieval method, as well as single-stage dense retrieval. During training, one can use dynamic retrieval of negatives through ANN search, following \cite{zhan_optimizing_2021}, although we defer investigating this for future work.
\end{itemize}
In summary, we can say that existing works modeling inter-document relationships and employing list-wise ranking loss functions are specific to the data modalities of ``deeply annotated'' datasets commonly used for list-wise ranking. As baselines they use models such as LambdaMART~\cite{Burges2010FromRT}, RankSVM~\cite{joachims_optimizing_2002} and GSF~\cite{ai_learning_2019}, and do not consider application of the proposed approaches to sparsely annotated, large scale deep learning datasets such as MS MARCO, nor do they compare or refer to any contemporary transformer-based dense retrieval or reranking models which currently represent the SOTA in ad hoc text retrieval. Likewise, the latter SOTA work does not consider list-wise ranking approaches, or refer and compare to the former line of work. Our present work aspires to bridge the gap between these two relatively isolated research communities by bringing list-wise, context-based ranking methodology to large, pre-trained transformer-based models, used in dense retrieval and trained through sparsely annotated, large scale datasets such as MS MARCO.

\subsection{Potential Risks}
The neural IR models in these studies, whether in their ``base'' form or extended by CODER, inherently contain societal biases and stereotypes. As discussed in previous studies~\cite{rekabsaz2021societal,rekabsaz2020neural}, these biases originate from already existing biases in the underlying transformer-based language models, as well as the fine-tuning process on the IR collections. Therefore, using these models in practice may lead to unfair treatment of various social groups (e.g. reflected in the representation or order of appearance in ranked lists of retrieval results), and we strongly advocate a conscious and responsible utilization of the models.

However, we note that CODER can potentially help alleviate this problem, as it offers a suitable framework for incorporating algorithmic bias mitigation methods into deep IR models. We have explored and verified this potential in follow-up work.

\subsection{Artifacts}

All artifacts used for this work (datasets, models, software) are publicly available and the terms of use are available on their respective sources. They have been used within their intended scope of use (research). Artifacts we created based on those may be restricted by terms and conditions specified by the original artifacts (derivatives of data accessed for research purposes should not be used outside of research
contexts).

\end{document}